# A mechanistic model of connector hubs, modularity, and cognition

Maxwell A. Bertolero[1,2*], B.T. Thomas Yeo[3,4], Danielle S. Bassett[2,5-7], Mark D'Esposito[1]

[1]Helen Wills Neuroscience Institute and the Department of Psychology, University of California, Berkeley, CA 94720-3190, USA
[2]Department of Bioengineering, School of Engineering and Applied Sciences, University of Pennsylvania, Philadelphia, PA 19104 USA
[3]Electrical and Computer Engineering, ASTAR-NUS Clinical Imaging Research Centre, Singapore Institute for Neurotechnology & Memory Networks Programme, National University of Singapore, Singapore 119077, Singapore
[4]NUS Graduate School for Integrative Sciences and Engineering, National University of Singapore, Singapore
[5]Department of Electrical & Systems Engineering, School of Engineering and Applied Sciences, University of Pennsylvania, Philadelphia, PA 19104, USA
[6]Department of Physics & Astronomy, College of Arts and Sciences, University of Pennsylvania, Philadelphia, PA 19104 USA
[7]Department of Neurology, Perelman School of Medicine, University of Pennsylvania, Philadelphia, PA 19104 USA.
*corresponding author (mbertolero@me.com)

## Abstract

The human brain network is modular—comprised of communities of tightly interconnected nodes[1]. This network contains local hubs, which have many connections within their own communities, and connector hubs, which have connections diversely distributed across communities[2,3]. A mechanistic understanding of these hubs and how they support cognition has not been demonstrated. Here, we leveraged individual differences in hub connectivity and cognition. We show that a model of hub connectivity accurately predicts the cognitive performance of 476 individuals in four distinct tasks. Moreover, there is a general optimal network structure for cognitive performance—individuals with diversely connected hubs and consequent modular brain networks exhibit increased cognitive performance, regardless of the task. Critically, we find evidence consistent with a mechanistic model in which connector hubs tune the connectivity of their neighbors to be more modular while allowing for task appropriate information integration across communities, which increases global modularity and cognitive performance.

## Main

The human brain is a complex network that can be parsimoniously summarized by a set of nodes representing brain regions and a set of edges representing the connections between brain regions. In network models of fMRI data, each edge represents the strength of functional connectivity—the temporal correlation of fMRI activity levels—between the two nodes (Figure 1). This network model can be used to study global and local brain connectivity patterns. Brain networks contain communities—groups of nodes that are more strongly connected to members of their own group than to members of other groups(Figure 1)[1,4,5]. This feature of networks is termed modularity and can be quantified by the modularity quality index $Q$ (see Methods for equation).

Modularity is ubiquitously observed in complex systems in Nature—a modular structure is observed consistently across the brains of very different species, from c elegans to humans[6]. Given its ubiquity, modular network organizations are potentially naturally selected because they reduce metabolic costs. Functional and structural connectivity is metabolically expensive[7-11]. A modular architecture with anatomically segregated and functionally specialized communities reduces the average length and number of connections—the network's wiring cost. Moreover, the brain's genotype-phenotype map is modular, forming groups of phenotypes, including brain communities (e.g., the visual community)[12,13], that are co-affected by groups of genes[14]. Modularity, at the genetic and phenotypic level, allows systems to quickly evolve under new selection pressures[15,16].

As we noted in earlier work[2], modularity potentially increases fitness in information processing systems [17-19], and network simulations show that modularity allows for robust network dynamics, in that the connections between nodes can be reconfigured without sacrificing information processing functions, a process necessary for the evolution of a network[20]. Artificial intelligence research has shown that modular networks also solve tasks faster and more accurately and evolve faster than non-modular networks [21] with lower wiring costs than non-modular networks[22]. Critically, modularity is also behaviorally relevant—modularity predicts intra-individual variation in working memory capacity[23] and how well an individual will respond to cognitive training[24,25].

Within each of these communities, local hubs exist that have strong connectivity to their own community. The within community strength can be used to measure a node's locality (see Methods for equation; Figure 1 for a schematic). A high within community strength reflects that a node has strong connectivity within its own community and is thus a local hub. Local hubs are ideally wired for segregated processing. Because the connections of local hubs are predominantly concentrated within their own community and their functions are likely specialized and segregated, damage to local hubs tends to cause relatively specific cognitive deficits[26,27] and does not significantly alter the modular organization of the network[27]. Supporting their more segregated and discrete role in information processing, their activity levels do not increase as more communities are involved in a task[1].

Yet, a completely modular organization renders the brain extremely limited in function— without connectivity between communities, information from, for example, visual cortex could never reach motor cortex and therefore visual information could not be used to inform movements. Thus, how is information integrated across these mostly segregated communities? The interdependence between modular communities and integration is a modern rendition of one of the first observations in neuroscience—Cajal's conservation principle, which states that the brain has been naturally selected and is thus organized by an economic trade-off between minimizing the wiring cost of the network, which leads to modularity, and more costly connectivity patterns that increase fitness, like the integrative functions afforded by connections between communities [28-30].

Connector hubs have diverse connectivity across different communities. The participation coefficient can be used to measure a node' diversity (see Methods for equation, Figure 1 for a schematic). A high participation coefficient reflects that a node has connections equally distributed across the brain's communities and is thus labeled a connector hub. Connector hubs are ideally wired for integrative processing[1,28,31-34]. In human brain networks, connector hubs have a particular cytoarchitecture[35], are implicated in a diverse range of cognitive tasks[36,37], and are physically located in anatomical areas at the boundaries between many communities[33]. Moreover, damage to connector hubs causes widespread cognitive deficits[26] and a decrease in the modular structure of the network[27]. During cognitive tasks, connector hubs appear to coordinate connectivity changes between other pairs of nodes—activity in connector hubs predicts changes in the connectivity of other nodes, particularly the connectivity between nodes in different communities[38-40]. Connector hubs are also strongly interconnected to each other, forming a diverse club—tightly interconnected connector hubs[5]. Connector hubs also have connections to almost every community in the network. Thus, they have access to information from every community. Finally, connector hubs exhibit increased activity if more communities are engaged in a task, which suggests that connector hubs are involved in processes that are more demanding as more communities are engaged[1].

Connector hubs might be Nature's cheapest solution to integration in a modular network. Generative models suggest that the diverse club—tightly interconnected connector hubs—potentially evolved to balance modularity and efficient integration[5]. However, given the amount of wiring required to link to many different and distant communities, connector hubs' connectivity pattern dramatically increases wiring costs[11]. Despite this cost, connector hubs potentially provide a necessary function—connector hubs could be the conductor of the brain's neural symphony.

A parsimonious mechanistic model of these findings is that connector hubs tune connectivity between communities. Neuronal tuning refers to cells selectively representing a particular stimulus, association, or information. We introduce the mechanistic concept of network tuning, in which connections between nodes are organized to achieve a particular network function or topology, like the integration of information across communities or decreased connectivity between two communities. We propose that diverse connectivity across the network's communities allows connector hubs to tune connectivity between communities to be modular but also allows for task appropriate information integration across communities. This facilitates a global modular network structure in which local hubs and nodes within each community are dedicated to mostly autonomous local processing. The modular network structure afforded by diversely connected connector hubs—connector hubs that are wired well for network tuning—is potentially optimal for a wide variety of cognitive processes. Thus, despite their cost, strong and diverse connector hubs might be critically necessary for integrative processing in complex modular neural networks.

Local and connector hubs have been exhaustively studied by network science and their functions have been inferred from their topological locations in the network[5]. Moreover, individuals' brain network connectivity has been shown to be predictive of task performance[41-45,46] and is able to "fingerprint" individuals[47]. However, no study has leveraged these individual differences to test a mechanistic model of hub function and direct evidence supporting a mechanistic model of these hubs and how they support human cognition remains absent. Moreover, it is currently unknown if there is a hub and network structure that is optimal for a diverse set of tasks or if different hub and network structures are optimal for different tasks. Here, we analyze how individual differences in the locality and diversity of hubs during the performance a task relates to network connectivity, modularity, and performance on that task as well as subject measures collected outside of the scanner, including psychometrics (e.g., fluid intelligence, working memory) and other behavioral measures (e.g., sleep quality and emotional states). We test a mechanistic model in which connector hubs tune their neighbors' connectivity to be more modular, which increases the global modular structure of the network and task performance, regardless of the particular task.

We leveraged the size and richness of fMRI data from 476 (S500 release) subjects that participated in the Human Connectome Project[48]. A network was built for each subject using fMRI data collected during seven different cognitive states (Resting-State, Working Memory, Social Cognition, Language & Math, Gambling, Relational, Motor, see Methods). Thus, each subject has seven different networks for analysis. Each edge represents the strength of functional connectivity between each pair of 264 nodes[49]. In every network, *Q* and a division of nodes into communities was calculated (see Methods). Next, in every network, for each node, locality and diversity was measured. Within community strength measures a node's locality and the participation coefficient measures a node's diversity, respectively (see Methods for equations). Figure 1 displays this processing workflow.

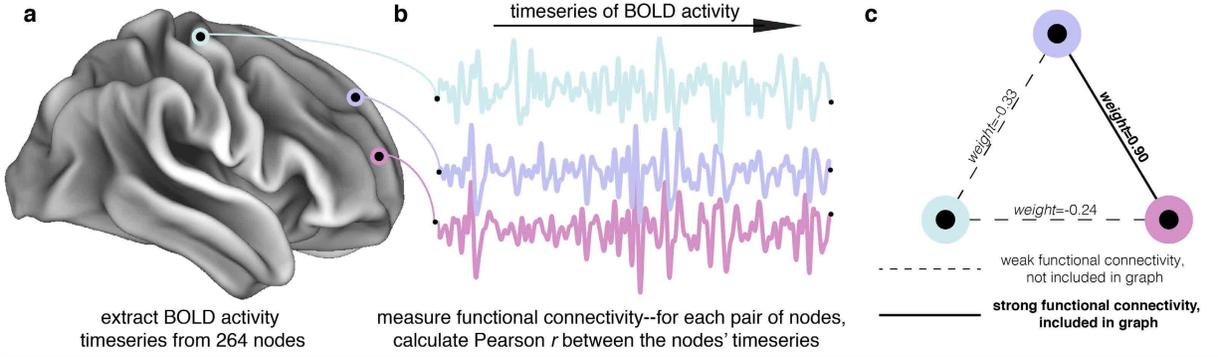

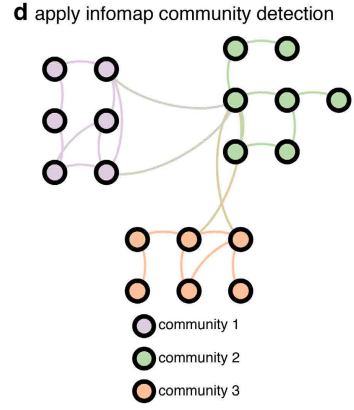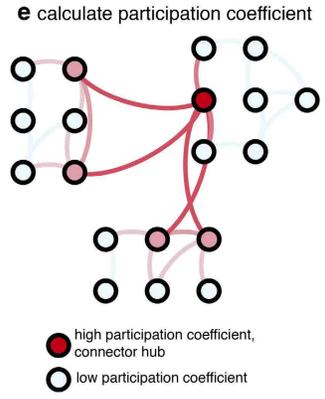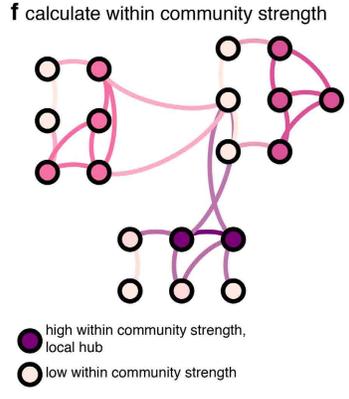

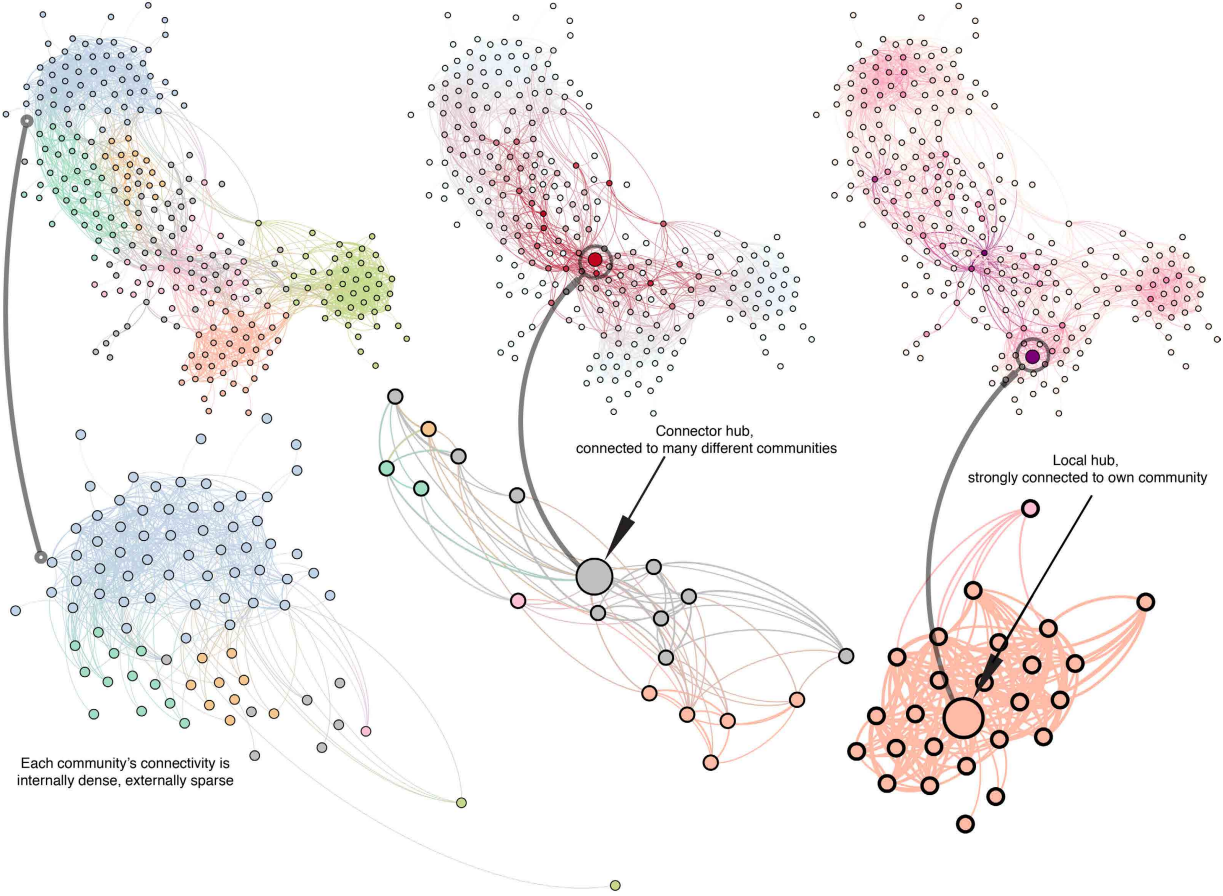

**Figure 1 | Functional connectivity and network science processing workflow. a**, The mean signal across time is extracted from 264 cortical, sub-cortical, and cerebellar regions, three of which are shown here. **b**, The time series of the three nodes is shown. To measure functional connectivity, the Pearson *r* correlation coefficient between the time series of node *i* and the time series of node *j* for all *i* and *j* is calculated. **c**, The strongest (e.g., the top 5% percent *r* values) functional connections serve as weighted edges in the graph (a range of graph densities was explored, see Methods for details). **d**, The Infomap community detection algorithm is applied, generating a community assignment for each node, displayed here in different colors in a schematic (top) and the mean graph across subjects (bottom). **e**, Given that particular community assignment and network, nodes' participation coefficients are calculated. Red nodes are high participation coefficient nodes, shown here in a schematic (top) and the mean graph (bottom). **f**, Within community strengths are also calculated. Purple nodes are high within community strength nodes, shown here in a schematic (top) and the mean graph (bottom). The graphs along the bottom are laid out using the force-atlas algorithm, where nodes are repelling magnets and edges are springs, which causes nodes in the same community to cluster together, nodes that are diversely connected across communities (connector hubs) to be in the center of the graph, and nodes that are strongly connected to a single community (local hubs) in the middle of that community. **d**, lower, A single community (light blue) and its connections to the rest of the graph is extracted and enlarged, with nodes colored by community. Note that the nodes within each community are more strongly connected to each other than to nodes in other communities. **e**, lower, A node (and its connections) with a high participation coefficient is extracted and enlarged, with nodes colored by community. Note that the connector hub is connected to many different communities. **f**, A node (and its connections) with a high within community strength is extracted and enlarged, with nodes colored by community. Note that the local hub is strongly connected to its own community.

In the proposed mechanistic model of hub connectivity, connector hubs, via their diverse connectivity, tune the network to preserve or increase global modularity and local hubs' locality, which, in turn, increases task performance. If this model is true, hubs' connectivity in the network and network modularity should be predictive of task performance. Thus, the first test of this model involved using hub diversity and locality, network connectivity, and modularity to predict task performance. A predictive multilayer perceptron model (three layers plus the input layer and the output layer (enough for non-linear relationships); eight neurons per layer (one per feature, with two layers containing 12 neurons, allowing for higher dimensional expansion)) was used to predict subjects' task performance (Supplementary Figure 1, Figure 2). Known as deep neural networks, these predictive models are constructed by tuning the weights between neurons across adjacent layers to achieve the most accurate relationship between the features (input) and the value the model is trying to predict (output). The predictive model's features ($n=8$) captured how well subjects' nodes' diversity and locality, network connectivity (i.e., edge weights in the network), and modularity ($Q$) are optimized for the performance of a task. For example, for the feature that captures how optimized the diversity of subjects' nodes' are for task performance, for each node, the Pearson *r* across subjects between that node's participation coefficients (which measures diversity) and task performance values was calculated (Supplementary Figure 1a). We call this *r* value the node's diversity facilitated performance coefficient. The feature, then, for a given subject, is the Pearson *r* across nodes between each node's diversity facilitated performance coefficient and each node's participation coefficient in that subject, representing how optimized the diversity of that subject's nodes' are for performance in the task (Supplementary Figure 1). Critically, for each subject's feature calculation, the diversity facilitated performance coefficients are calculated without that subject's data. The same procedure is executed for locality (using the within community strengths) and edge weights; instead of participation coefficients, within community strengths or edge weights are used. Finally, the $Q$ values of the networks are included in the model.

The predictive model was fit for each of the four cognitive tasks that subjects performed in the Human Connectome Project for which performance was measured (Working

Memory, Relational, Language and Math, Social tasks; see Methods for task performance measures). For this and other subject performance analyses, we could not analyze the Gambling, Motor, or Resting-State tasks, as there was no performance measured for these tasks. Each predictive model was fit to the subjects' networks constructed during the performance of each task as well as the resting state (four features from each). The inclusion of the resting-state and the cognitive task state allowed the model to capture the subjects' so-called intrinsic network states as well as the subjects' task driven network states. Using a leave-one-out cross-validation procedure, the features were constructed, and the model was fit with data from all subjects except one. The predictive model was then used to predict the left-out subject's task performance (Supplementary Figure 1c). To test the accuracy of the model, the Pearson $r$ between the observed and predictive performance of each subject was calculated (Figure 2, Supplementary Figure 2).

This predictive model significantly ($p<0.001$, Bonferroni corrected ($n$ tests = 4)) predicted performance in all four tasks (Figure 2). Also, using a predictive model with only nodes' diversity and locality and modularity features (i.e., ignoring individual connections in the network) did not dramatically decrease the models' prediction accuracies (Supplementary Figure 2a,b). Given that head motion is a concern when analyzing fMRI data, scrubbing techniques were applied to remove motion artifacts from the fMRI data and the mean frame-wise displacement was regressed out from task performance. Neither of these additional analyses dramatically decreased the predictive models' prediction accuracies (Supplementary Figure 2c-f). Finally, in each task, modularity ($Q$) alone was only weakly correlated with task performance (Working Memory, Pearson's $r$ (dof=471):0.303, $p<0.001$, CI:0.219,0.383; Relational, Pearson's $r$ (dof=455):0.106, $p$:0.095, CI:0.014,0.196; Language & Math, Pearson's $r$ (dof=469):0.085, $p$:0.259, CI:-0.005,0.174; Social, Pearson's $r$ (dof=471):0.084, $p$:0.275, CI:-0.006,0.173, all confidence intervals=95%). These results suggest that the diversity and locality of nodes, in combination with the modular connectivity structure of the network, are highly predictive of task performance.

The Human Connectome project contains psychometrics and other behavioral measures collected outside of the MRI scanner; for clarity and to differentiate these measures from the task performance measures and the tasks' corresponding networks, we call these "subject measures" [50]. If a particular hub and network structure is generally optimal for many different types of cognition and many different behaviors (a component of the mechanistic model of hub function), then the tasks' optimal hub and network structures should be similarly optimal across subject measures—sub-optimal for negative measures like poor sleep, sadness, and anger and optimal for positive measures like life satisfaction and processing speed.

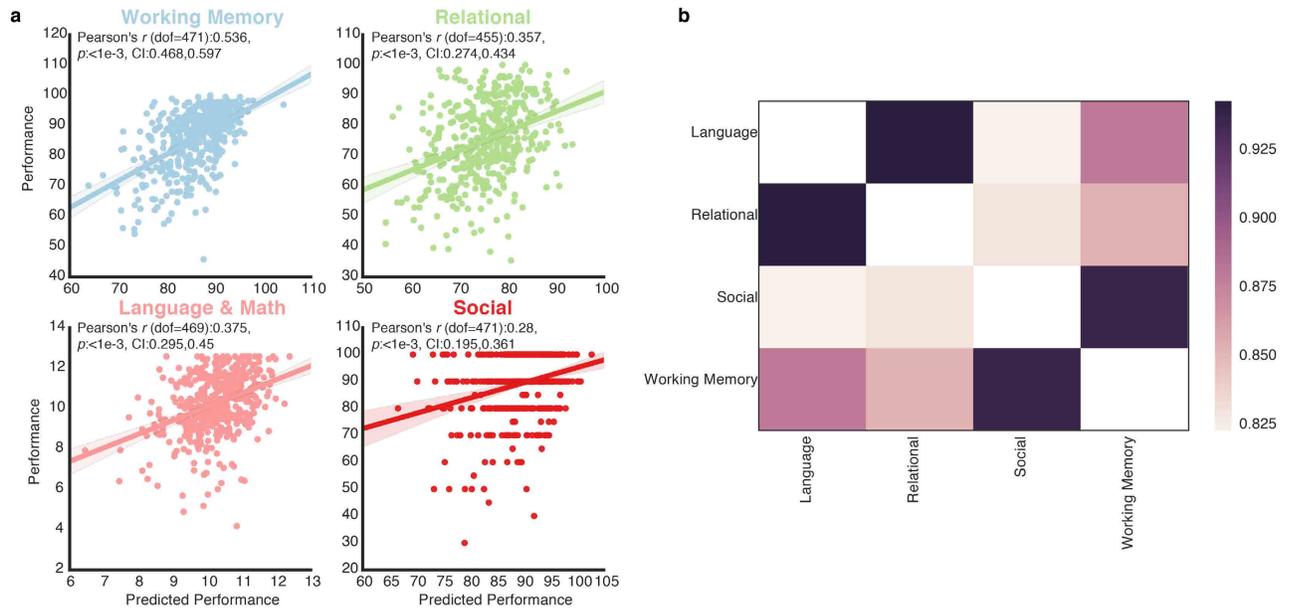

**Figure 2 | Hub diversity and locality, modularity, and network connectivity predict cognitive performance**. **a**, for each task, the correlation between task performance and the performance predicted by a predictive model of hub diversity and locality, modularity, and network connectivity. Each data point represents the (y-axis) true performance (see Methods, each task's performance value scale is unique) of the subject and the (x-axis) predicted performance of the subject by the neural network. Shaded areas represent 95 percent confidence intervals. In every task, the predictive model significantly predicted task performance ($p < $ 1e-3, Bonferroni corrected ($n$ tests=4), N=Working Memory: 473, Relational: 457, Language & Math: 471, Social: 473). **b**, we correlated the tasks' feature correspondence values (see Supplementary Figure 3 for each task's feature correspondence with each subject measure)—measuring if the two tasks' optimal hub and network structures are also optimal for the same subject measures. High correlations mean that the two tasks' hub and network structures are similarly optimal for the same subject measures (all results significant at $p < $ 1e-3, Bonferroni corrected ($n$ tests = 4), dof=45, $N$=47, the number of subject measures, while the feature correspondence $N$ =Working Memory: 473, Relational: 457, Language & Math: 471, Social: 473).

For each task, the predictive model constructs features that capture how optimal each subject's hub and network structure is for performance on that task. Using the subjects' networks from a given task, the predictive model of hub and network structure can construct features that capture how optimal each subject's hub and network structure is for a given subject measure collected outside of the scanner instead of task performance. This was executed using the networks from each of the four tasks for all subject measures. Thus, the predictive model constructs features that capture how optimal each subject's hub and network structure (measured during the performance of a task (e.g. Working Memory)) is for a given subject measure (e.g., Delayed Discounting). The correspondence between the features in the two models—how similarly optimal subjects' hub and network structure are for the task and a given subject measure—can then by calculated by, across subjects, computing the correlation between the features in the two predictive models. Specifically, for each feature, the correlation, across subjects, between the feature in the predictive model fit to task performance (e.g., Working Memory) and that feature in the predictive model that was fit to a subject measure (e.g., Delayed Discounting) is computed. The mean correlation across the edge, locality, and diversity features ($n$=6, three features from resting-state and three features from the task) is then calculated, which we call feature correspondence. The $Q$ feature was ignored, as the $Q$ feature remains constant regardless of what the model is fit to. Thus, this value determines if each task's optimal hub and network structure is optimal for other subject measures and if all the tasks' optimal hub and network structures are similarly optimal for other subject measures.

For each task, the hub and network structures that were optimal for that task were typically also optimal for positive subject measures and sub-optimal for negative subject measures (Supplementary Figure 3). Next, the similarity by which two tasks' optimal hub and network structures generalized to other subject measures can be measured by correlating the feature correspondence values (for example, the Working Memory and Social columns in Supplementary Figure 3). High Pearson $r$ correlations were found between all tasks ($r$ (dof=45) values between 0.82 and 0.96, $p$<0.001 Bonferroni corrected ($n$ tests = 4), Figure 2b). Finally, the predictive model was able to significantly predict most subject measures (Supplementary Figure 4). These results demonstrate that, if an individual has a particular brain state during a given task, as defined by the connectivity of the network's hubs, that is optimal for that given task, it also likely optimal for other subject measures. Critically, different tasks' optimal hub and network structures are similarly optimal for other subject measures. Moreover, these findings demonstrate that the predictive model captures hub connectivity patterns in the network that are relevant for behavior and cognition in general, instead of overfitting hub connectivity patterns that are only related to a particular cognitive process or behavior.

Having established relationships between hub locality and diversity, modularity, and task performance, we sought to test the mechanistic claim that diverse connector hubs increase modularity by analyzing how individual differences in a node's diversity within the network are predictive of individual differences in brain network modularity ($Q$; see Methods for mathematical definition). Typically, the result of damage to a region can be used to infer the function of that region—if a region is damaged and modularity decreases, the region is putatively involved in preserving modularity. Here, we analyze the other direction—when a hub is diversely connected across the brain (i.e., strong), if modularity increases, the region's diverse connectivity is putatively involved in preserving modularity (Supplementary Figure 5).

Thus, we first tested if, across subjects, a node's participation coefficients are positively correlated with modularity ($Q$). For each node, the Pearson $r$ between that node's participation coefficients and the network's modularity values ($Q$) across subjects was calculated. Intuitively, higher $r$ values indicate that the node's diversity (i.e., the participation coefficient) is associated with higher network modularity. This is an important feature that can be used to distinguish the roles of different brain regions. For ease of presentation, we refer to each node's $r$ value as the diversity facilitated modularity coefficient, as it measures how the diversity of the node's connections facilitates (we use this term to remain causally agnostic) the modularity of the network. For every node, the Pearson $r$ between the within community strengths and $Q$ values across subjects was also calculated. Intuitively, higher $r$ values indicate that the node's locality (i.e., the within community strength) is associated with higher network modularity. We refer to each node's $r$ value (between within community strengths and $Q$ values across subjects) as the locality facilitated modularity coefficient, as it measures how the locality of the node's connections facilitates the modularity of the network.

We performed these computations separately for all seven distinct cognitive states. In all states, the diversity facilitated modularity coefficients of connector hubs (top 20 percent highest participation coefficient nodes) were shown to be significantly higher than other nodes in a Bonferroni-corrected independent two-tailed *t*-test (Figure 3a, Working Memory *t*(dof:262):7.182, *p*<0.001, Cohen's *d*:1.104, CI:0.062,0.117, Gambling *t*(dof:262):4.101, *p*:0.0004, Cohen's *d*:0.63, CI:0.025,0.052, Language & Math *t*(dof:262):7.292, *p*<0.001, Cohen's *d*:1.12, CI:0.062,0.102, Motor *t*(dof:262):7.354, *p*<0.001, Cohen's *d*:1.13, CI:0.088,0.13, Relational *t*(dof:262):4.457, *p*:0.0001, Cohen's *d*:0.685, CI:0.038,0.075, Resting State *t*(dof:262):3.947, *p*:0.0007, Cohen's *d*:0.606, CI:0.029,0.096, Social *t*(dof:262):3.716, *p*:0.0017, Cohen's *d*:0.571, CI:0.022,0.051. *P* values Bonferroni corrected (*n* tests=7), all confidence intervals=95%). Moreover, in all cognitive states, the locality facilitated modularity coefficients of local hubs (top 20 percent highest within community strength nodes) were shown to be significantly higher than other nodes in a Bonferroni-corrected independent two-tailed *t*-test (Figure 3b, Working Memory *t*(dof:262):5.415, *p*<0.001, Cohen's *d*:0.832, CI:0.045,0.093, Gambling *t*(dof:262):4.959, *p*<0.001, Cohen's *d*:0.762, CI:0.034,0.074, Language & Math *t*(dof:262):6.428, *p*<0.001, Cohen's *d*:0.988, CI:0.045,0.085, Motor *t*(dof:262):9.822, *p*<0.001, Cohen's *d*:1.509, CI:0.101,0.146, Relational *t*(dof:262):6.131, *p*<0.001, Cohen's *d*:0.942, CI:0.036,0.07, Resting State *t*(dof:262):0.966, *p*:1.0, Cohen's *d*:0.148, CI:-0.014,0.038, Social *t*(dof:262):4.54, *p*:0.0001, Cohen's *d*:0.698, CI:0.026,0.06. *P* values Bonferroni corrected (*n* tests = 7), all confidence intervals=95%). While the diversity facilitated modularity coefficients of connector hubs were not always positive, they were typically close to or above zero. This means that a diverse connector hub can be associated with increased integrative connectivity between communities without decreasing the modularity of the network. To more fully understand the relationship between nodes' diversity and the networks' modularity, the Pearson *r* between each node's mean participation coefficient across subjects (which defines a connector hub) and the node's diversity facilitated modularity coefficient was calculated (Supplementary Figure 6). Moreover, the Pearson *r* between each node's mean within community strength across subjects (which defines a local hub) and the node's locality facilitated modularity coefficient was calculated (Supplementary Figure 6). In every task, there was a significant positive correlation between a node's mean participation coefficient and that node's diversity facilitated modularity coefficient (Supplementary Figure 6a). In every task, there was also a significant positive correlation between a node's mean within community strength and its locality facilitated modularity coefficient (Supplementary Figure 6b). These analyses demonstrate that connector hubs' strong diverse connectivity to many communities and local hubs' strong local connectivity is associated with higher brain network modularity, regardless of the subjects' cognitive state. Thus, these results are consistent with the mechanistic model of connector hub function, where connector hubs preserve the modular structure of the network via diverse connectivity.

We confirmed the reliability and reproducibility of these results and demonstrated that they are not driven by analytically necessary relationships. First, the mean participation coefficient and within community strength was calculated in one half of the subjects and the diversity and locality facilitated modularity coefficients were calculated in the other half of the subjects, testing 10,000 splits (Supplementary Figure 7). Next, four null models were tested to ensure the current results were not driven by analytically necessary relationships (Supplementary Figure 8). Other analyses ensured the current results are not driven by the number of communities (Supplementary Figure 9, Supplementary Figure 10). Finally, to justify the use of the Pearson $r$ to calculate the coefficients, the relationship between nodes' diversity and $Q$ was confirmed as typically linear (Supplementary Figure 11).

Having found evidence supporting a mechanistic model in which connector hubs tune their neighbors' connectivity to be more modular, thereby increasing the global modular structure of the network, we next asked if diverse hubs concurrently facilitate higher modularity and higher task performance. To address this question, the Pearson $r$ between each node's participation coefficient (or within community strength) and task performance was calculated. A positive $r$ at a node indicates that a subject with a higher participation coefficient (or within community strength) at that node performs better on the task. We refer this $r$ value as the node's diversity facilitated performance coefficient for participation coefficients, and locality facilitated performance coefficient for within community strengths. Note that these are the same $r$ values used in the construction of the predictive performance model features (Supplementary Figure 1). However, for the predictive model, we calculated these $r$ values with the subject whose behavior was to be predicted held out. Here, we calculate these $r$ values across all subjects.

In all tasks, the diversity facilitated performance coefficients of connector hubs (top 20 percent strongest) were shown to be significantly higher than other nodes in a Bonferroni-corrected independent two-tailed $t$-test, with only the Language & Math task at $p=0.0677$ after Bonferroni correction (uncorrected $p=0.0169$) (Figure 3c, Working Memory $t$(dof:262):5.378, $p<0.001$, Cohen's $d$:0.826, CI:0.03,0.071, Language & Math $t$(dof:262):2.404, $p$:0.0677, Cohen's $d$:0.369, CI:0.005,0.04; Relational $t$(dof:262):2.959, $p$:0.0135, Cohen's $d$:0.455, CI:0.01,0.037; Social $t$(dof:262):4.744, $p<0.001$, Cohen's $d$:0.729, CI:0.025,0.053. All $p$ values Bonferroni corrected ($n$ tests=4), all confidence intervals=95%). The locality facilitated performance coefficients of local hubs (top 20 percent strongest) were shown to be significantly higher than other nodes in a Bonferroni-corrected independent two-tailed $t$-test (Figure 3d, Working Memory $t$(dof:262):2.712, $p$:0.0285, Cohen's $d$:0.417, CI:0.008,0.054, Language & Math $t$(dof:262):2.864, $p$:0.0181, Cohen's $d$:0.44, CI:0.006,0.043; Relational $t$(dof:262):0.327, $p$:1.0, Cohen's $d$:0.05, CI:-0.016,0.021; Social $t$(dof:262):1.862, $p$:0.2547, Cohen's $d$:0.286, CI:0.0,0.031. All $p$ values Bonferroni corrected ($n$ tests=4), all confidence intervals=95%). Moreover, the correlation between each node's diversity facilitated performance coefficient and each node's mean participation coefficient was positive and significant (Supplementary Figure 6), suggesting that, for connector hubs, a higher participation coefficient is associated with higher task performance. The Pearson $r$ correlation between each node's locality facilitated performance coefficient and each node's mean within community strength was also positive and significant (Supplementary Figure 6; all tasks except Relational, Bonferroni ($n$ tests=4) $p$=0.081, uncorrected $p$=0.02), suggesting that, for local hubs, a higher within community strength is also associated with higher task performance. Finally, there was a significant positive correlation between a node's diversity facilitated modularity coefficient and a node's diversity facilitated performance coefficient (Figure 3e) as well as a significant positive correlation between a node's locality facilitated modularity coefficient and a node's locality facilitated performance coefficient (Figure 3f). Thus, diverse connector hubs facilitate higher task performance in proportion to how much they facilitate higher modularity, suggesting a strong link between the increased modularity afforded by diverse hubs and increased task performance.

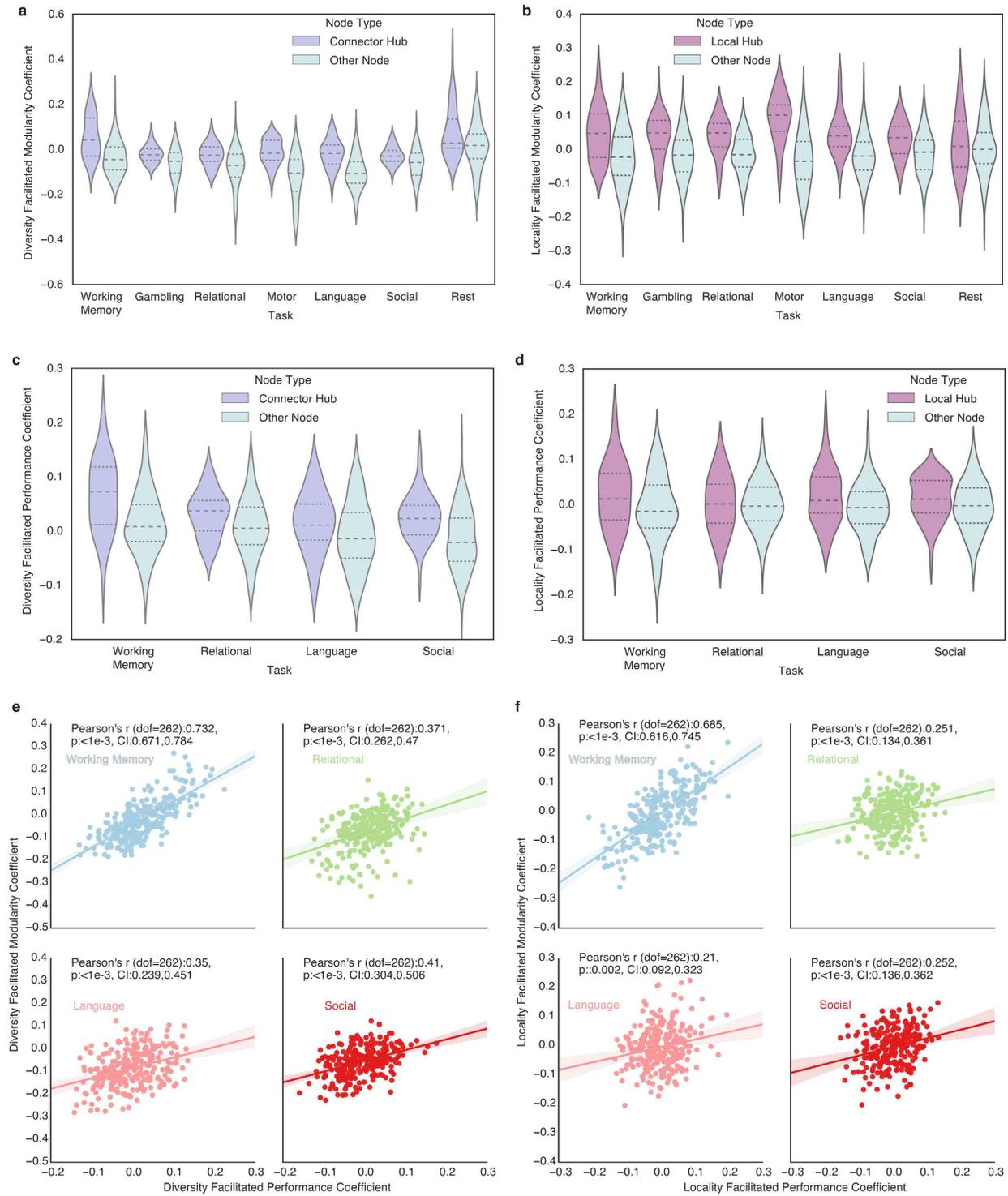

**Figure 3 | Connector hubs and local hubs concurrently facilitate increased modularity and task performance**. For each task, diversity and locality facilitated modularity coefficients, a measure of how the diversity and locality (respectively) of a node facilitates modularity, were calculated. In every task, the diversity and locality facilitated modularity coefficients of connector (**a**) and local hubs (**b**), compared to other nodes, is significantly (except Resting-State for locality) higher, demonstrating that strong connector and local hubs facilitate the modular structure of brain networks. For each task, diversity and locality facilitated performance coefficients were calculated. In every task the diversity and locality facilitated performance coefficients of connector (**c**) and local hubs (**d**), compared to other nodes, is significantly (except Language for diversity ($p=0.0677$ after Bonferroni correction (uncorrected $p=0.0169$)), Relational and Social for locality) higher, demonstrating that strong connector and local hubs facilitate increased task

performance. For **a-d**, the mean and quartiles are marked in each violin. Each task's distribution of coefficients was tested for normality using D'Agostino and Pearson's omnibus test $k^2$. No evidence was found ($k^2$>0.0 for all tasks) that these distributions were not normal. *N*=264, the number of nodes in the graph. **e**, The correlation between a node's diversity facilitated modularity coefficient and a node's diversity facilitated performance coefficient. **f**, The correlation between the node's locality facilitated modularity coefficient and the node's locality facilitated performance coefficient. In panels **e,f**, *N*=264, the number of nodes in the graph. Shaded areas represent 95 percent confidence intervals. All *p* values are Bonferroni corrected (*n* tests = 4).

Next, we tested the mechanistic network tuning claim of the model: "do connector hubs increase *Q* by tuning the connectivity of their neighbors' edges to be more modular?" This relationship should only hold for connector hubs, not local hubs, as previous studies suggest that connector hubs tune connectivity between communities and maintain a modular structure[1,36,38-40]. We therefore examined connector hubs for which their diversity facilitated modularity coefficients were positive. This analysis had two aspects. First, do connector hubs increase modularity by tuning within community edge strengths? Second, are connector hubs tuning the within community edge strengths of their neighbors in order to increase global modularity?

In order to test the first aspect of the neural tuning mechanism—if within community edges are tuned by connector hubs in order to increase global modularity—we used a canonical division of nodes into communities (Figure 4d displays this division, see Methods for link to division[36,49]). We assessed, for each edge in the network, how the edge's weights related to modularity values (*Q*) across subjects (Figure 4a). Next, we calculated how well each connector hub's participation coefficients correlate with each edge's weights across subjects (Figure 4b). Higher *Q* values and higher connector hub participation coefficients are associated with decreased connectivity between the visual, sensory/motor hand, sensory/motor mouth, auditory, ventral attention, dorsal attention, and cingulo-opercular communities. These communities were also more strongly connected to fronto-parietal, default mode, salience, and sub-cortical communities in networks with higher modularity values and higher connector hub participation coefficients.

Given these observations, we sought to find the edges that mediate between connector hubs' increased participation coefficients and modularity (*Q*), as these are the edges that connector hubs likely tune in order to increase *Q*. Specifically, a mediation analysis was performed for each connector hub, with an edge weight mediating the relationship between the connector hub's participation coefficients and the *Q* indices of the networks across subjects. An edge's mediation value of a connector hub's participation coefficients and *Q* is the regression coefficient of the edge's weights by the connector hub's participation coefficients across subjects multiplied by the regression coefficient of *Q* indices by the edge's weights, controlling for the connector hub's participation coefficients, across subjects. Each edge's mean mediation value across connector hubs is shown in Figure 4c. We found that edges between the visual, sensory/motor hand, sensory/motor mouth, auditory, ventral attention, dorsal attention, and cingulo-opercular communities, as well as edges between those communities and the fronto-parietal, default mode, and sub-cortical communities, mediate the relationship between connector hubs' participation coefficients and *Q* indices. These results are consistent with a mechanistic model in which diverse connector hubs tune connectivity to increase segregation between sensory, motor, and attention systems, which increases the global modularity of the network.

Next, we tested the second aspect of the network tuning mechanism—if the relationship between a connector hub's participation coefficients and $Q$ indices is mediated primarily by that connector hub's neighbors' edge pattern increasing $Q$. Neighbors were defined based on edges present between the two nodes in a graph at a density of 0.15 (as it was our densest cost explored). The mediation values calculated above each represent an edge mediating between a node $i$'s participation coefficients and $Q$ values. Thus, for each connector hub $i$, there is the set of arrays of absolute mediation values of node $i$'s neighbors' edges ($n=263$ for each neighbor $j$'s array) and the set of arrays of the absolute mediation values of node $i$'s non-neighbors' edges ($n=263$ for each non-neighbor $j$'s array). Edges of node $i$ in every array were ignored, as we were only interested in how the participation coefficients of connector hub $i$ modulate $Q$ via the mediation of $j$'s connectivity to the rest of the network, not $j$'s connectivity to connector hub $i$ (thus, $n=264-1$). If a connector hub is primarily modulating $Q$ via the tuning of its neighbors' edges, then the absolute mediation values in the neighbors' arrays should be greater than the absolute values in the non-neighbors' arrays. The distribution of $t$-values between the two sets of arrays for all connector hubs (neighbors versus non-neighbors) is shown in Supplementary Figure 12; across tasks, the mediation values were consistently and significantly higher for connector hubs' neighbors' edges than non-neighbors edges. Moreover, these same $t$-values can be calculated for local hubs, using the within community strength instead of the participation coefficient; thus, an edge mediates between a local hub's within community strengths and $Q$. Across tasks, connector hubs' neighbors' mediation $t$ values were shown to be higher than local hubs' neighbors' mediation $t$ values with a two-tailed independent student's $t$-test ($t$(dof:1358): 3.892, $p$:0.0001, Cohen's $d$:0.219, CI:1.887,6.62), demonstrating that this result is specific to connector hubs. All distributions were confirmed as normal ($k^2>100.0$, $p<0.00001$ for all tasks). We also performed an alternative analysis that confirmed these relationships (see Methods, Supplementary Figure 13). These results suggest that each connector hub, not local hub, tunes their neighbors' connectivity to be more modular. A connector hub's high diversity facilitated modularity coefficient does not largely reflect diffuse global connectivity changes. Instead, connector hubs are likely connected in a way that allows them to directly tune the connectivity of their neighbors to be more modular, thereby increasing global network modularity. Thus, the locality facilitated modularity coefficients are likely a downstream effect of connector hub modulation. Supporting this interpretation, we found that, when connector hubs have high participation coefficients, local hubs have high within community strengths (Supplementary Figure 14).

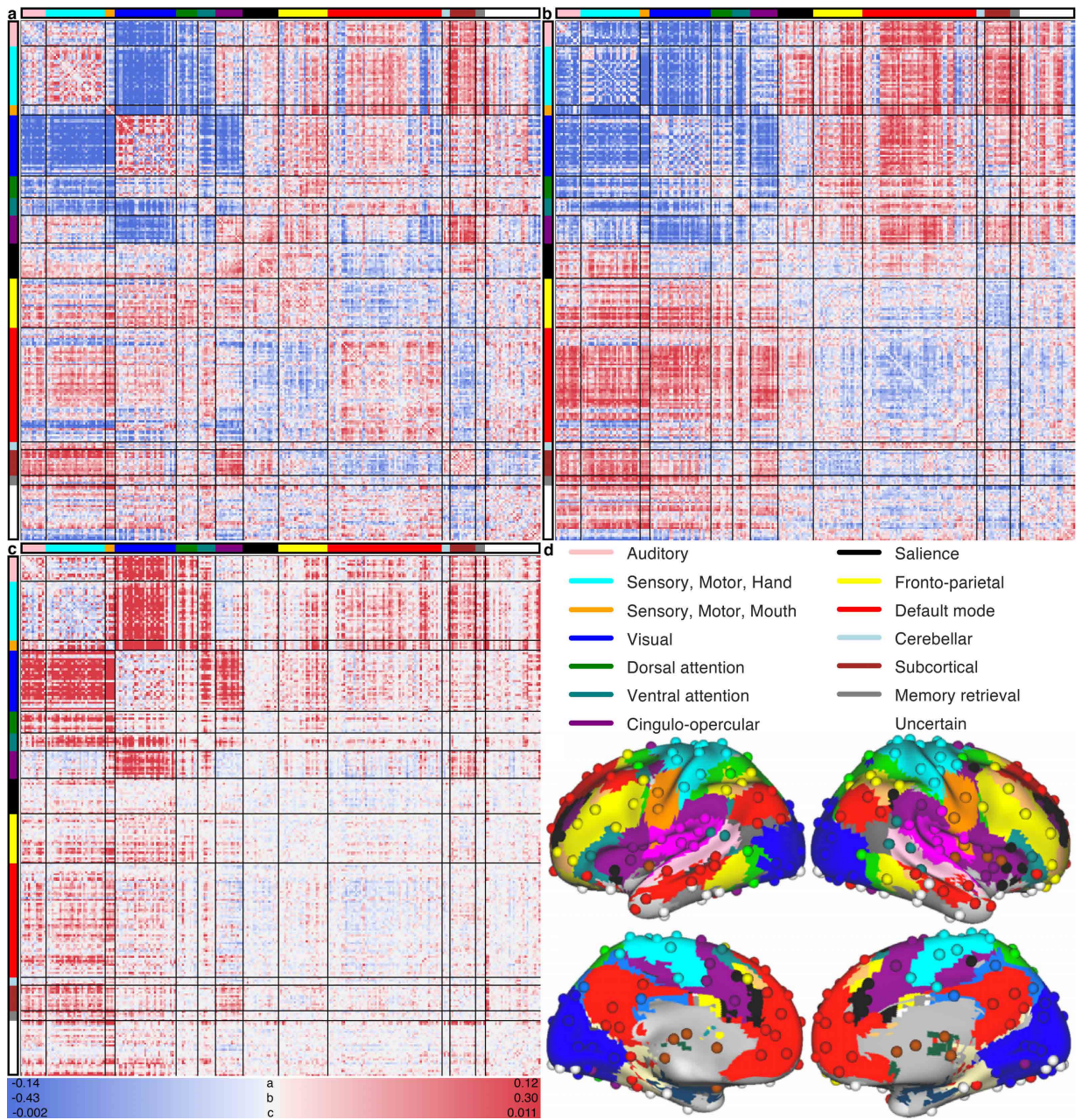

**Figure 4 | Connectivity between primary sensory, motor, dorsal attention, ventral attention, and cingulo-opercular communities mediate the relationship between connector hubs and modularity**. **a**, Each entry is the Pearson correlation coefficient, $r$, across subjects ($N$=476), between modularity ($Q$) and that edge's weights. **b**, For each connector hub, the Pearson $r$ between the hub's participation coefficients and each edge's weights across subjects ($N$=476) was calculated. The matrix in **b** is the mean of those matrices across connector hubs. **c**, To investigate the relationship between connector hubs' participation coefficients, edge weights, and $Q$, a mediation analysis was performed for each connector hub, with an edge's weights mediating the relationship between the connector hub's participation coefficients and $Q$ indices ($N$=476). Each edge's mean mediation value between connector hubs' participation coefficients and $Q$ is shown. **d**, The anatomical locations of each node and community on the cortical surface[36,49].

In the series of analyses we report here, we explicitly and comprehensively tested a mechanistic model by leveraging individual differences in connectivity and cognition in humans. Specifically, a model of the diversity and locality of hubs, the modularity of the network, and the network's connectivity was highly predictive of task performance and a range of subject measures. Critically, the diversity and locality of nodes optimal for each task were also similarly optimal for positive subject measures. Thus, it appears that there is a hub and network structure that is generally optimal for cognitive processing. We found evidence that diverse connector hubs preserve or increase the modularity of brain networks. Moreover, diverse connector hubs tune the connectivity of their neighbors to be more modular. Finally, we found that the diversity of connector hubs simultaneously facilitated higher modularity and task performance. Thus, connector hubs appear to contribute to the maintenance of an optimal modular architecture during integrative cognition without greatly increasing the wiring cost or decreasing modularity[5,29,51]. In sum, these data are consistent with a mechanistic model of hub function, where connector hubs integrate information and subsequently tune their neighbors' connectivity to be more modular, which increases the global modularity of the network, allowing local hubs and nodes to perform segregated processing.

Across individuals, we found that diverse connector hubs increase modularity and task performance, regardless of the task. In all seven tasks, the subjects with the most diversely connected connector hubs also had the highest modularity and, in the four tasks for which performance was measured, the highest task performance. Thus, we propose that connector hubs are likely critical for integrating information and tuning their neighbors' connectivity to be more modular, regardless of the task. Although connector hubs are more active during tasks that require many communities[1], as their functions are likely more computationally demanding during these tasks, it is likely that every task requires the functions of connector hubs, as supported by our finding that their diverse connectivity predicts performance in all of the tasks analyzed here.

Our findings compliment many previous task-based fMRI studies that have identified regions that are more active during a particular cognitive process. We have demonstrated that, while different regions are more or less active in different tasks, including connector hubs, the diverse connectivity and integration and tuning functions of connector hubs are consistently required across different cognitive processes. Our findings are also consistent with neuropsychological studies of patients with focal brain lesions. It has been found that damage to connector hubs decreases modularity and causes widespread cognitive deficits, while damage to local nodes does not decrease modularity and causes more isolated deficits, such as hemiplegia, or aphasia[26,27]. While connector hubs are not likely critical for only one specific cognitive process, their functions and diverse connectivity are required to maintain a cognitively optimal modular structure across cognitive processes. Thus, as we observed here, individual differences in the diversity of connector hubs' connectivity is predictive of cognitive performance across a range of very different tasks. Although diversely connected connector hubs are critical for successful performance in many different tasks, any given task nevertheless recruits very different cognitive and neural processes; each task likely engages connectivity patterns that are specifically optimal to that task. Future analyses should seek to understand both the general optimal connectivity patterns of connector hubs found here and the connectivity patterns that are optimal to a single task, including if and how these connectivity patterns interact.

## Methods

**Data and Preprocessing** We used fMRI data from the Human Connectome Project[48] S500 release. For the task-based fMRI data, Analysis of Functional NeuroImages (AFNI) was used to preprocess the images[52]. The AFNI command *3dTproject* was used, passing the mean signal from the cerebral spinal fluid mask, the white matter mask, the whole brain signal, and the motion parameters to the "-ort" options, which removes these signals via linear regression. Within AFNI, the "-automask" option was used to generate the masks. The "-passband 0.009 0.08" option, which removes frequencies outside of 0.009 and 0.08, was used. Finally, the "-blur 6" option, which smooths the images (inside the mask only) with a 6mm FWHM filter after the time series filtering. Given the short length of the Emotion task (176 frames; Resting-State:1200, Social: 274, Relational:232, Motor:284, Language:316, Working Memory:405, Gambling:253) it was not included in our analyses. For the fMRI data collected at rest, we used the images that were previously preprocessed by the Human Connectome Project with ICA-FIX. We also used the AFNI command *3dBandpass* to further preprocess these images. We used it to remove the mean whole brain signal and frequencies outside 0.009 and 0.08 (explicitly, "-ort whole_brain_signal.1D -band 0.009 0.08 -automask"). We did not regress out stimulus or task effects from the time series of each node, because how nodes' low frequency oscillations respond to stimulus or task effects is meaningful. Moreover, other investigators have noted that task effect regression has minimal effects[53].

As subject head motion during fMRI can impact functional connectivity estimates and has been shown to bias brain-task performance relationships[54], performance prediction analyses were executed with scrubbing (removing frames with high motion) executed on frames with frame-wise displacement greater than 0.2 millimeters, including the frame before and after the movement. Frame-wise displacement measures movement of the head from one volume to the next, and was computed as the sum of the absolute values of the differentiated rigid body realignment estimates (translation and rotation in x, y, and z directions) at every time point with rotation values evaluated with a radius of 50 mm[54]. Frames were removed after all preprocessing was executed. Subjects with more than 75 percent of frames removed were not analyzed. Moreover, we executed all analyses after regressing out mean frame-wise displacement from the task performance values (Supplementary Figure 2).

**Graph Theory Analyses** The Power atlas[49] was used to define the 264 nodes in our graph because it was the only atlas that met all of the following requirements: (1) Given that the homogeneity of nodes in this atlas is high and they do not share physical boundaries, it will not overestimate the local connectivity of regions, (2) it is the only atlas that is defined based both on functional connectivity and studies of task activations making it optimal for our current analyses, (3) it accurately divides nodes into communities observed with other approaches (e.g., at the voxel level), and this division has been used in many studies[33,36,49,55]. A canonical division of nodes into communities aides in the interpretation and generalizability of our results. It can be found at: http://www.nil.wustl.edu/labs/petersen/Resources_files/Consensus264.xls. Moreover, we used this division to calculate within and between community edge weight changes across subjects. (4) It has anatomical coverage of cortical, subcortical, and cerebellar regions.

All graph theory analyses were executed with our own custom python code (www.github.com/mb3152/brain_graphs) that uses the iGraph library. All analysis code is also publicly available (github.com/mb3152/hcp_performance/). For each task (both LR and RL encoding directions were used) and for each subject, the mean signal from 264 regions in the Power atlas was computed. The Pearson $r$ between all pairs of signals was computed to form a 264 by 264 matrix, which was then Fisher z transformed. We chose Pearson $r$ values to represent functional connectivity (i.e., edges) between nodes, for its simplicity in interpretation and ubiquity in human network neuroscience[56]. However, more complex statistical measures could be employed, including measures that attempt to estimate the directionality of each edge. The LR and RL matrices were then averaged. The mean matrix was then thresholded, retaining edge weights, at a range of costs (0.05 to 0.15 at 0.01 intervals), a common range and interval in graph theory analyses[1,27,33,49]. The maximum spanning tree was calculated to ensure all nodes had at least one edge. No negative correlations were included in our analyses. The matrix was then normalized to sum to a common value across subjects, and was used to represent the edges in the graph. Thus, all graphs had the same number of edges and sum of edge weights.

For each cost, the InfoMap algorithm[57] was run. While this method has been shown to be highly accurate on benchmark networks with known community structures, it is still a heuristic, as community detection is NP-hard[58]. While InfoMap does not explicitly maximize $Q$, it has been shown to estimate community structure accurately in several test cases[59], rendering the $Q$ value, the participation coefficients, and within community strengths computed based on the community structure accurate and valid. Moreover, in biological networks, InfoMap achieves $Q$ values that are similar to algorithms that maximize $Q$ [60]; in the current resting-state data, InfoMap $Q$ values and Fast-Greedy $Q$[61] values were correlated at Pearson $r$=0.87 (dof=474, $p$<0.001, 95% CI: 0.84, 0.89); InfoMap $Q$ values were found to be higher than Fast-Greedy $Q$ values with a student's independent $t$-test (t(dof:952):16.027, $p$:<0.001, Cohen's d:0.775, 95% CI:0.024,0.03). InfoMap $Q$ values and Louvain $Q$ [62] values were correlated at Pearson $r$=0.98 (dof=474, p<0.001, 95% CI: (0.97, 0.98)); Louvain $Q$ values were higher than InfoMap $Q$ values. When comparing InfoMap $Q$ values to the distribution that includes both Louvain and Fast-Greedy, two algorithms that explicitly maximize $Q$, InfoMap $Q$ values were shown to be significantly higher with a student's independent $t$-test ($t$(dof:1429):5.304, $p$:<0.001, Cohen's d:0.222, 95% CI:0.005,0.011). Regardless, we found that it detects a community structure with $Q$ values highly similar to other methods (Supplementary

Figure 15). Moreover, previous work has demonstrated the stability of community detection and the participation coefficient across community detection methods[5].

The participation coefficients, within community strengths, and $Q$ were calculated at each cost. $Q$ is written analytically as follows. Consider a weighted and undirected graph with $n$ nodes and $m$ edges represented by an adjacency matrix $\mathbf{A}$ with elements

$$\mathbf{A}_{ij} = \text{edge weight between i and j.}$$

Thus, the strength of a node is given by

$$k_i = \sum_j \mathbf{A}_{ij}$$

And modularity ($Q$) can be written as:

$$Q = \frac{1}{2m} \sum_{i \neq j} (\mathbf{A}_{ij} - \gamma p_{ij}) \delta(c_i, c_j).$$

Here, $p_{ij}$ is the probability that nodes $i$ and $j$ are connected in a random null network

$$P_{ij} = \frac{k_i k_j}{2m},$$

$\gamma$ is the resolution parameter, and $c_i$ is the community to which node $i$ belongs to and $\delta(\alpha, \beta) = 1$ if $\alpha = \beta$ and $\delta(\alpha, \beta) = 0$ if $\alpha \neq \beta$.

Given a particular community assignment, the participation coefficient of each node can be calculated. The participation coefficient (PC) of node $i$ is defined as:

$$PC_i = 1 - \sum_{s=1}^{N_M} \left(\frac{K_{is}}{K_i}\right)^2$$

where $K_i$ is the sum of $i$'s edge weights, $K_{is}$ is the sum of $i$'s edge weights to community $s$, and $N_M$ is the total number of communities. Thus, the participation coefficient is a measure of how evenly distributed a node's edges are across communities. A node's participation coefficient is maximal if it has an equal sum of edge weights to each community in the network. A node's participation coefficient is 0 if all of its edges are to a single community.

Finally, we calculate the within community strength value for each node as follows:

$$z_i = \frac{k_i - \bar{k}_{s_i}}{\sigma_{k_{s_i}}}$$

Where $k_i$ is the number of links of node $i$ to other nodes in its community $s_i$, $\bar{k}_{s_i}$ is the average of $k$ over all the nodes in $s_i$, and $\sigma_{k_{s_i}}$ is the standard deviation of $k$ in $s_i$. Thus, The within community strength measures how well-connected node $i$ is to other nodes in the community relative to other nodes in the community.

Each subject's participation coefficient, within community strength, and $Q$ were the mean of those values across the range of costs. All analyses were executed and all prediction models were fit separately for each task.

**Tasks** The following descriptions for each task have been adapted for brevity from the Human Connectome Project Manual[63].

Working Memory. The category specific representation task and the working memory task are combined into a single task paradigm. Participants were presented with blocks of trials that consisted of pictures of places, tools, faces and body parts (non-mutilated parts of bodies with no "nudity"). Within each run, the 4 different stimulus types were presented in separate blocks. Also, within each run, 1⁄2 of the blocks use a 2-back working memory task and 1⁄2 use a 0-back working memory task (as a working memory comparison). A 2.5 second cue indicates the task type (and target for 0-back) at the start of the block. Each of the two runs contains 8 task blocks (10 trials of 2.5 seconds each, for 25 seconds) and 4 fixation blocks (15 seconds). On each trial, the stimulus is presented for 2 seconds, followed by a 500 ms inter-task interval (ITI).

Gambling. Participants play a card guessing game where they are asked to guess the number on a mystery card (represented by a "?") in order to win or lose money. Participants are told that potential card numbers range from 1-9 and to indicate if they think the mystery card number is more or less than 5 by pressing one of two buttons on the response box. Feedback is the number on the card (generated by the program as a function of whether the trial was a reward, loss or neutral trial) and either: 1) a green up arrow with "$1" for reward trials, 2) a red down arrow next to -$0.50 for loss trials; or 3) the number 5 and a gray double headed arrow for neutral trials. The "?" is presented for up to 1500 ms (if the participant responds before 1500 ms, a fixation cross is displayed for the remaining time), following by feedback for 1000 ms. There is a 1000 ms ITI with a "+" presented on the screen. The task is presented in blocks of 8 trials that are either mostly reward (6 reward trials pseudo randomly interleaved with either 1 neutral and 1 loss trial, 2 neutral trials, or 2 loss trials) or mostly loss (6 loss trials pseudo- randomly interleaved with either 1 neutral and 1 reward trial, 2 neutral trials, or 2 reward trials). In each of the two runs, there are 2 mostly reward and 2 mostly loss blocks, interleaved with 4 fixation blocks (15 seconds each).

Motor. Participants are presented with visual cues that ask them to either tap their left or right fingers, or squeeze their left or right toes, or move their tongue to map motor areas. Each block of a movement type lasted 12 seconds (10 movements), and is preceded by a 3 second cue. In each of the two runs, there are 13 blocks, with 2 of tongue movements, 4 of hand movements (2 right and 2 left), and 4 of foot movements (2 right and 2 left). In addition, there are 3 15-second fixation blocks per run.

Language & Math. The task consists of two runs that each interleave 4 blocks of a story task and 4 blocks of a math task. The lengths of the blocks vary (average of approximately 30 seconds), but the task was designed so that the math task blocks match the length of the story task blocks, with some additional math trials at the end of the task to complete the 3.8 minute run as needed. The story blocks present participants with brief auditory stories (5-9 sentences) adapted from Aesop's fables, followed by a 2-alternative forced- choice question that asks participants about the topic of the story. For example: *after a story about an eagle that saves a man who had done him a favor, participants were asked, "Was that about revenge or reciprocity?"* The math task also presents trials auditorily and requires subjects to complete addition and subtraction problems. The trials present subjects with a series of arithmetic operations (e.g., "fourteen plus twelve"), followed by "equals" and then two choices (e.g., "twenty- nine or twenty- six"). Participants push a button to select either the first or the second answer. The tasks are adaptive to try to maintain a similar level of difficulty across participants.

Social (Theory of Mind). Participants were presented with short video clips (20 seconds) of objects (squares, circles, triangles) that either interacted in some way, or moved randomly on the screen. After each video clip, participants judge whether the objects had a mental interaction (an interaction that appears as if the shapes are taking into account each other's feelings and thoughts), Not Sure, or No interaction (i.e., there is no obvious interaction between the shapes and the movement appears random). Each of the two task runs has 5 video blocks (2 Mental and 3 Random in one run, 3 Mental and 2 Random in the other run) and 5 fixation blocks (15 seconds each).

Relational. The stimuli are 6 different shapes filled with 1 of 6 different textures. In the relational processing condition, participants are presented with 2 pairs of objects, with one pair at the top of the screen and the other pair at the bottom of the screen. They are told that they should first decide what dimension differs across the top pair of objects (differed in shape or differed in texture) and then they should decide whether the bottom pair of objects also differ along that same dimension (e.g., if the top pair differs in shape, does the bottom pair also differ in shape). In the control matching condition, participants are shown two objects at the top of the screen and one object at the bottom of the screen, and a word in the middle of the screen (either "shape" or "texture"). They are told to decide whether the bottom object matches either of the top two objects on that dimension (e.g., if the word is "shape", is the bottom object the same shape as either of the top two objects. For both conditions, the subject responds yes or no using one button or another. For the relational condition, the stimuli are presented for 3500 ms, with a 500 ms ITI, and there are four trials per block. In the matching condition, stimuli are presented for 2800 ms, with a 400 ms ITI, and there are 5 trials per block. Each type of block (relational or matching) lasts a total of 18 seconds. In each of the two runs of this task, there are 3 relational blocks, 3 matching blocks and 3 16-second fixation blocks.

**Performance measures**. All performance measures were chosen *a priori*. In the working memory task, we used the mean accuracy across all n-back conditions (face, body, place, tool). In the relational task, we used mean accuracy across both the matching and the relational conditions. For the language task, we took the maximum difficulty level that the subject achieved across both the math and language conditions. We did not use accuracy, because the task varies in difficulty based on how well the subject is doing, making accuracy an inaccurate measure of performance for these tasks. For the social task, given that almost all subjects correctly identified the social interactions as social interactions, we used the percentage of correctly identified random interactions.

**Deep neural network model**. A deep neural network is a supervised learning algorithm that can learn a non-linear function for regression or classification. Unlike logistic regression, there are one or more non-linear layers, called hidden layers, between the input and the output layer. Thus, the model is trained to relate a set of input features to outputs by learning weights between neurons across adjacent layers (Supplementary Figure 1). Our implementation uses the sklearn python library. Explicitly, a prediction for subject z is calculated as:

```
model = sklearn.neural_network.MLPRegressor(hidden_layer_sizes=(8,12,8,12))
model.fit(x[subjects!=z], y[subjects!=z])
prediction = model.predict(x([z])
```

where *x* is the set of features across subjects and *y* is the task performance across subjects.

**Analytic quality of diversity and locality facilitated modularity coefficients** To further understand diversity and locality facilitated modularity coefficients, we performed an iterative split-half analysis. Specifically, we estimated the mean within community strength or participation coefficient of each node in one half of subjects, and each node's locality and diversity facilitated coefficient in the other half, testing 10,000 random splits of subjects. All relationships were reliably observed in every cognitive state (Supplementary Figure 7). Next, we sought to determine if this relationship was a necessary feature of the underlying mathematics, or whether it was a phenomenon specific to the neurophysiology of brain networks. To address this question, we tested four null model networks and observed that none of them exhibited a significant relationship between mean participation coefficient and diversity facilitated modularity coefficient (Supplementary Figure 8). As a third check, we assessed whether the number of communities identified in the network was inadvertently biasing our results. We observed that the number of communities in each network was negatively correlated with the modularity value *Q* (Supplementary Figure 9). After regressing out the number of communities in each network from the modularity value, we observed that our findings remained qualitatively unchanged (Supplementary Figure 10). Finally, we tested whether the relationships between variables of interest were linear (and therefore appropriate to examine with Pearson *r* correlation coefficients), or nonlinear. To address this question, we analyzed individual 1$^{st}$, 2$^{nd}$, and 3$^{rd}$ order curve fits of the relationship between participation coefficients and modularity values. We observed that many relationships were well-captured by a first order fit, with the connector hub's maximal participation coefficients corresponding to maximal *Q* indices, with only a few showing a more nonlinear relationship (Supplementary Figure 11).

**Alternative analysis of connector hubs' tuning the connectivity of their neighbors**
We executed an alternative analysis to test if connector hubs tune the connectivity of their neighbors to be more modular. For each node *i* we calculated a matrix, where the *j-k*$^{th}$ entry is the Pearson *r* correlation coefficient that captures how well the participation coefficients of node *i* correlates with the edge weights between nodes *j* and *k* in the network across subjects. These Pearson *r* values allowed us to test whether a node's participation coefficients correlate positively with its neighbors' increased connectivity to its own community and decreased connectivity to other communities. We subtracted the sum of *r* values in the matrix corresponding to node *i*'s participation coefficients and node *j*'s between community edge weights from the sum of *r* values in the matrix corresponding to node *i*'s participation coefficients and node *j*'s within community edge weights. Thus, this value measures how well the participation coefficients of node *i* are correlated with the increased modular (within community) connectivity of node *j*. We used the partition of nodes into communities that was created along with the nodes themselves (Figure 6d)[49]. Edges between node *i* and node *j* were ignored in this calculation, as the participation coefficients of node *i* is likely highly correlated with the edge weights between node *i* and node *j*, and we were only interested in how the participation coefficient of node *i* modulates node *j*'s connectivity to the rest of the network, not node *j*'s connectivity to node *i*. Edges that were not positive on average across subjects were not included in this analysis, as the interpretation of negative edges in fMRI-based networks is not obvious (results were similar only including the top 25 percent of edges (Supplementary Figure 13). Correlations between the edge strength between nodes *i* and *j* and the amount of modulation of *j*'s modularity by *i* were calculated such that the set of nodes *i* were either connector hubs or non-connector hubs. A positive correlation means that a node is biased to modulate the connectivity of its neighbors versus its non-neighbors to be more modular. In all cognitive states, these correlations were only positive and significant (Pearson's *r*>0.17, *p*<0.001, Bonferroni corrected (*n* tests = 7)) for connector hubs (Supplementary Figure 13) suggesting that connector hubs tune the connectivity of their neighbors to be more modular.

To test if this relationship existed for local hubs, we calculated a similar matrix, where, for each node *i*, the *j-k* $^{th}$ entry is the Pearson *r* value that captures how well the within community strengths of node *i* correlate with the edge weight between nodes *j* and *k* in the network across subjects. Correlations between the edge strength between nodes *i* and $j$ and the amount of modulation of $j$'s modularity by the within community strength of $i$ were calculated such that nodes $i$ were either local hubs or non-local hubs. None of these correlations were robust (-0.1>$r$<0.1). These analyses add to our conclusion, demonstrated in the Results, that connector hubs facilitate higher modularity by tuning the connectivity of their neighbors to be more modular.

**Statistical Methods**

The number of subjects was determined by the number of subjects released by the Human Connectome Project at the start of the analyses. As this dataset represented the largest dataset of its kind at that time and the number of subjects is greater than many similar analyses[47], no power analysis was computed. Total $N$=Working Memory: 475, Gambling: 473, Relational:458, Motor:475, Language & Math:472, Social:474, Resting State: 476. However, as we only analyzed subjects with both Resting-State and the task scans, $N$=Working Memory:473, Relational:457, Language & Math:471, Social:473. This results in a unique $N$=476 across tasks, in that 476 different subjects had a resting state scan and at least one task scan. As scrubbing (which removes frames with large head motion) can cause too many frames to be removed from the time series, subjects with less than 75 percent of remaining frames were not included in the analyses that implemented scrubbing; thus, for analyses using scrubbed data, $N$=Working Memory:351, Relational:335, Language & Math:348, Social:358.

All confidence intervals (CI) are reported with alpha=0.05. For Pearson $r$ correlation coefficients CIs, the interval of $r$ values is given by Fisher transforming $r$ to $z$, computing the interval, and then Fisher reverse transforming the $z$ intervals back to $r$ intervals. For t-tests, the confidence interval represents the largest and smallest differences in means across the two distributions. For all t-tests, distributions were confirmed as normal ($p<0.001$) or exhibiting no significant evidence as not normal ($k^2>0.0$) using D'Agostino and Pearson's omnibus test $k^2$. All $p$ values are two sided tests.

All $p$ values that are part of a family of tests are Bonferroni corrected for multiple comparisons. For example, we test if two tasks' hub and network structures are similarly optimal for the same subject measures, testing across a large number of subject measures. In this case, we applied a Bonferroni correction to the $p$-values to determine whether the effect remained true for particular subject measures. Here, the number of tests is equal to the number of subject measures, 47. Individual subject networks were built independently for each task and task performance is different for each task. Thus, these tests are not strictly in the same family. However, to be conservative, we still Bonferroni corrected these $p$-values. In these cases, the family size is either 4 or 7, depending on the number of tasks analyzed. Unless otherwise stated, all $p$ values are Bonferroni corrected.

Many statistical tests are calculated here without reported $p$ values. For example, Pearson $r$ values are used to calculate functional connectivity. Here, only the $r$ values are of interest—more precisely, individual differences in the $r$ values across subjects, and how these differences relate to individual differences in cognition. This treatment of multiple comparisons in the context of functional connectivity and individual differences in cognition is common and recommended[47,64]. We extend this notion to other analyses here as well. For example, we use the Pearson correlation coefficient $r$ to compare how well different nodes' participation coefficients across subjects explain variance in network modularity or task performance (the diversity facilitated modularity and performance coefficients). In these cases, we relate these $r$-values to other measures, and are only concerned with how these $r$-values explain another distributions' variance (here, we find a positive correlation between these $r$-values and a node's mean participation coefficient across subjects). We are not concerned with the statistical significance any particular $r$-value as estimated by the $p$-value. We care about the distribution of $r$-values, not the distribution of $p$-values, and we do not make any claims about any single $r$-value. Thus, the $p$-values are neither reported nor corrected for multiple comparisons. This is precisely how functional connectivity is treated statistically.

**Acknowledgements** This work was supported by NIH Grant NS79698 and the National Science Foundation Graduate Research Fellowship Program under Grant no. DGE 1106400 to MAB and MD. MAB would also like to acknowledge NIH T32 Ruth L. Kirschstein Institutional National Research Service Award (5T32MH106442-02). BTTY was also supported Singapore MOE Tier 2 (MOE2014-T2-2-016), NUS Strategic Research (DPRT/944/09/14), NUS SOM Aspiration Fund (R185000271720), Singapore NMRC (CBRG/0088/2015), NUS YIA and the Singapore National Research Foundation (NRF) Fellowship (Class of 2017). DSB would also like to acknowledge support from the John D. and Catherine T. MacArthur Foundation, the Alfred P. Sloan Foundation, the Army Research Laboratory and the Army Research Office through contract numbers W911NF-10-2-0022 and W911NF-14-1-0679, the National Institute of Health (2-R01-DC-009209-11,1R01HD086888-01, R01-MH107235, R01-MH107703, and R21-M MH-106799), the Office of Naval Research, and the National Science Foundation (BCS-1441502, PHY-1554488, and BCS-1631550). The funders had no role in study design, data collection and analysis, decision to publish, or preparation of the manuscript.

**Data availability**
Data were provided by the Human Connectome Project, WU-Minn Consortium (Principal Investigators: David Van Essen and Kamil Ugurbil; 1U54MH091657) funded by the 16 NIH Institutes and Centers that support the NIH Blueprint for Neuroscience Research and by the McDonnell Center for Systems Neuroscience at Washington University. The content is solely the responsibility of the authors and does not necessarily represent the official views of any of the funding agencies. All analyses were executed in accordance with the authors' institutions' relevant ethical regulations as well as the WU-Minn HCP Consortium Open Access Data Use Terms. Informed consent was obtained from all participants.

**Competing Interests**
The authors declare no competing interests.

**Code availability**
All graph theory analyses were executed with our own custom python code (www.github.com/mb3152/brain_graphs) that uses the iGraph library. All analysis code is also publicly available (github.com/mb3152/hcp_performance/).

**Author Contributions** M.A.B. conceived the analyses. M.A.B., B.T.T.Y., D.S.B., and M.D. collaboratively designed the analyses. M.A.B. executed the analyses. M.A.B., B.T.T.Y., D.S.B., and M.D. collaboratively wrote the paper.

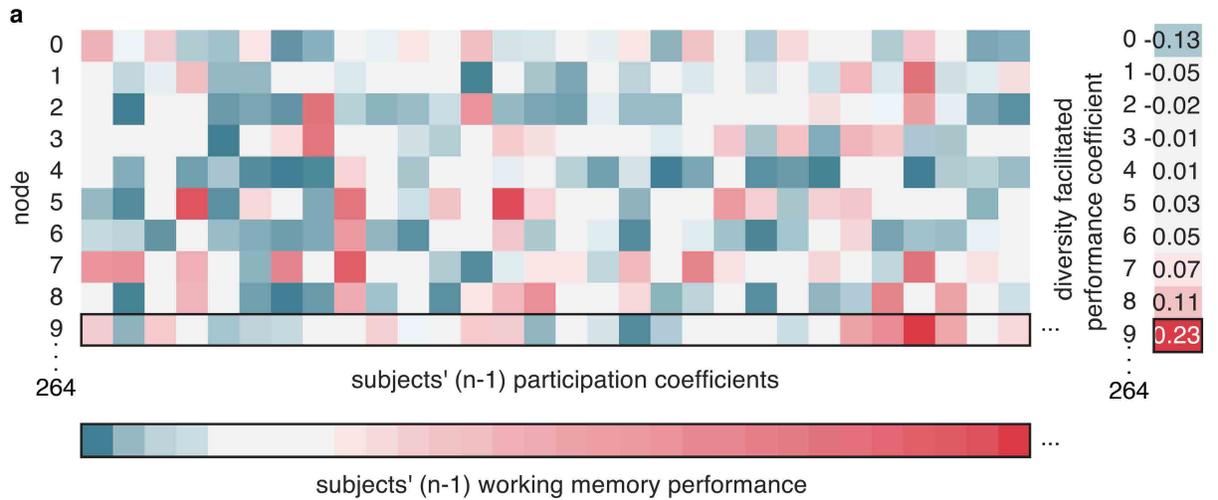

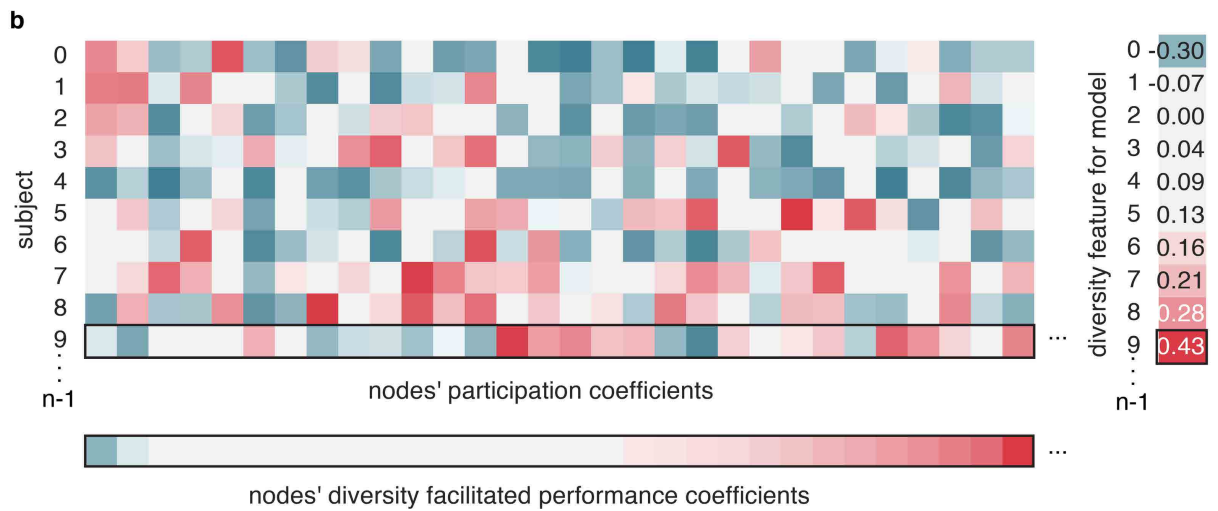

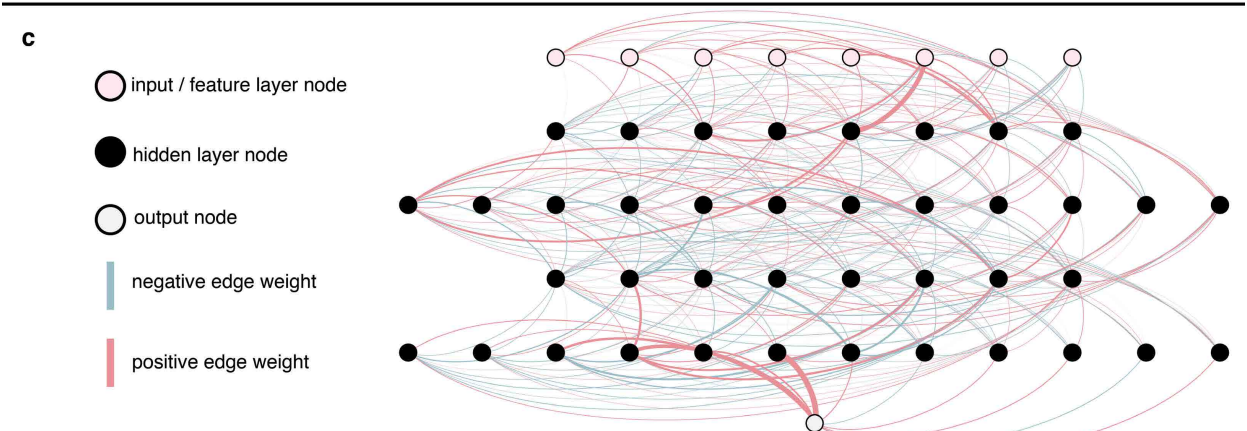

Supplementary Figure 1. **Workflow of the deep neural network features and model construction. a**, For each node, we measured the diversity facilitated performance coefficient. For example, the Pearson *r* is calculated between node 9's participation coefficients across subjects (black outline) and task performance across subjects (black outline), resulting in a diversity facilitated performance coefficient of 0.23 (black outline) for that node. **b**, The optimality of a subject's participation coefficients for task performance is measured by, for example, calculating the Pearson *r* between subject 9's participation coefficients (black outline) and the diversity facilitated performance coefficients (black outline), resulting in a diversity feature for the model of 0.43 (black outline).

The same procedures in **a** and **b** are also executed using within community strengths and edge weights instead of participation coefficients. Modularity (Q) values are also used as features. These four features are derived separately from resting-state and the task network for which task performance is being predicted, resulting in eight features. All calculations in panels **a** and **b** are calculated without data from the subject for which the prediction is being made. These eight features are then used in a deep neural network model (**c**) to learn the relationship between the features and task performance (again, without data from the subject for which the prediction is being made) by adjusting the weights between nodes in adjacent layers. The features are then calculated for the left-out subject (e.g., Pearson *r* between the left-out subject's participation coefficients and the previously calculated diversity facilitated performance coefficients that did not include data from the left-out subject) and are used in the deep neural network model to generate a prediction for the left-out subject.

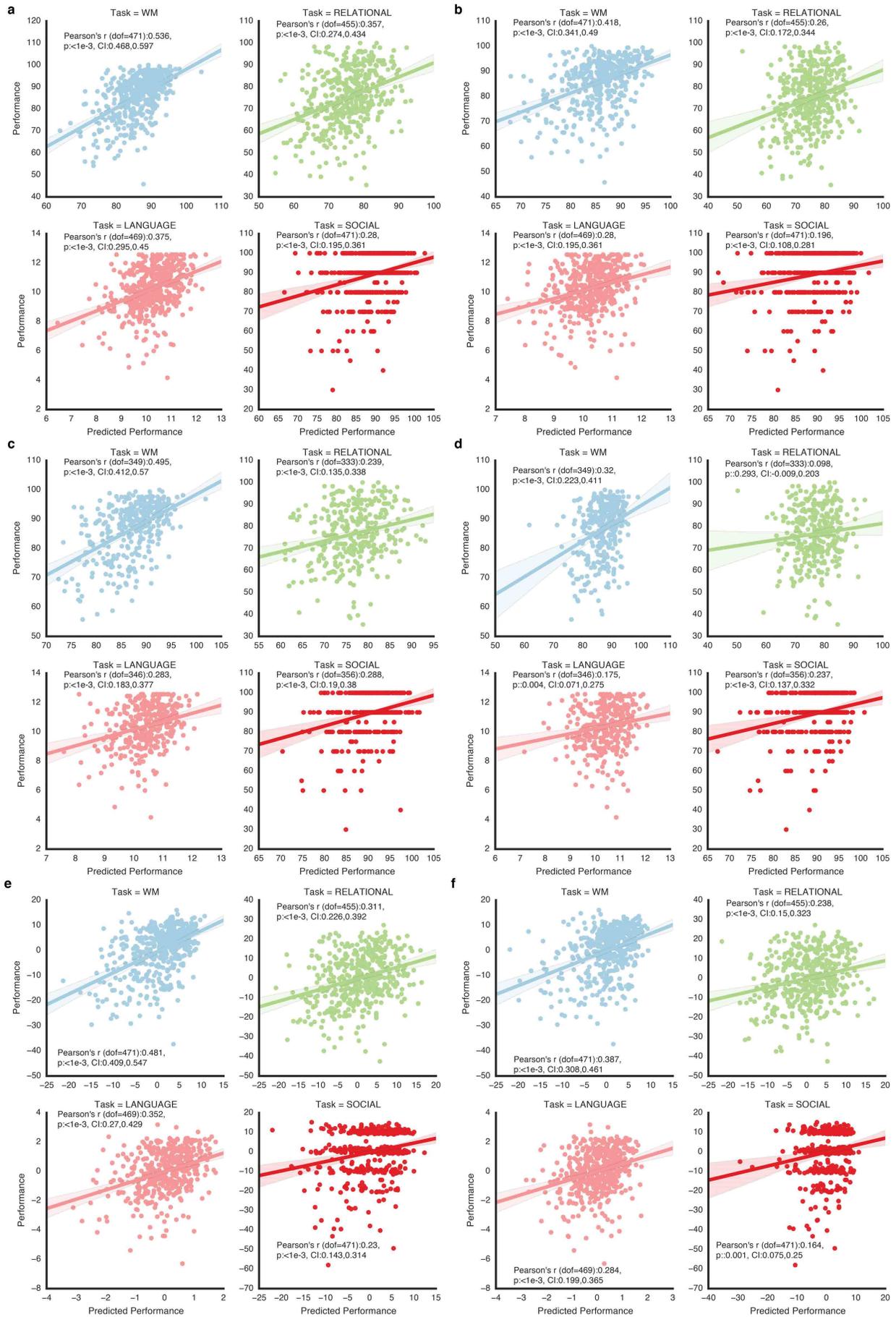

Supplementary Figure 2 | **Hub diversity and locality, modularity, and network connectivity predicts task performance**. **a**, predictive model of task performance (see Methods for performance measures), as shown in Figure 2a. **b**, Predictive model of task performance without using network connectivity (i.e., only hub diversity and locality and modularity). *N*=Working Memory: 473, Relational: 457,Language & Math: 471, Social: 473). Each dot represents a subject's prediction. Shaded areas represent 95 percent confidence intervals. **c,d**, As in **a,b**, except the frames with high motion were removed from the time series before the network of that individual was constructed (i.e., scrubbing), *N*=Working Memory: 351, Relational: 335, Language & Math: 348, Social: 358, as subjects with less than 75 percent of frames after scrubbing were not analyzed. **e,f**, As in **a,b**, except mean frame wise displacement was regressed out from the task performance values, *N*=Working Memory: 473, Relational: 457,Language & Math: 471, Social: 473). All *p* values are Bonferroni corrected (*n* tests=4).

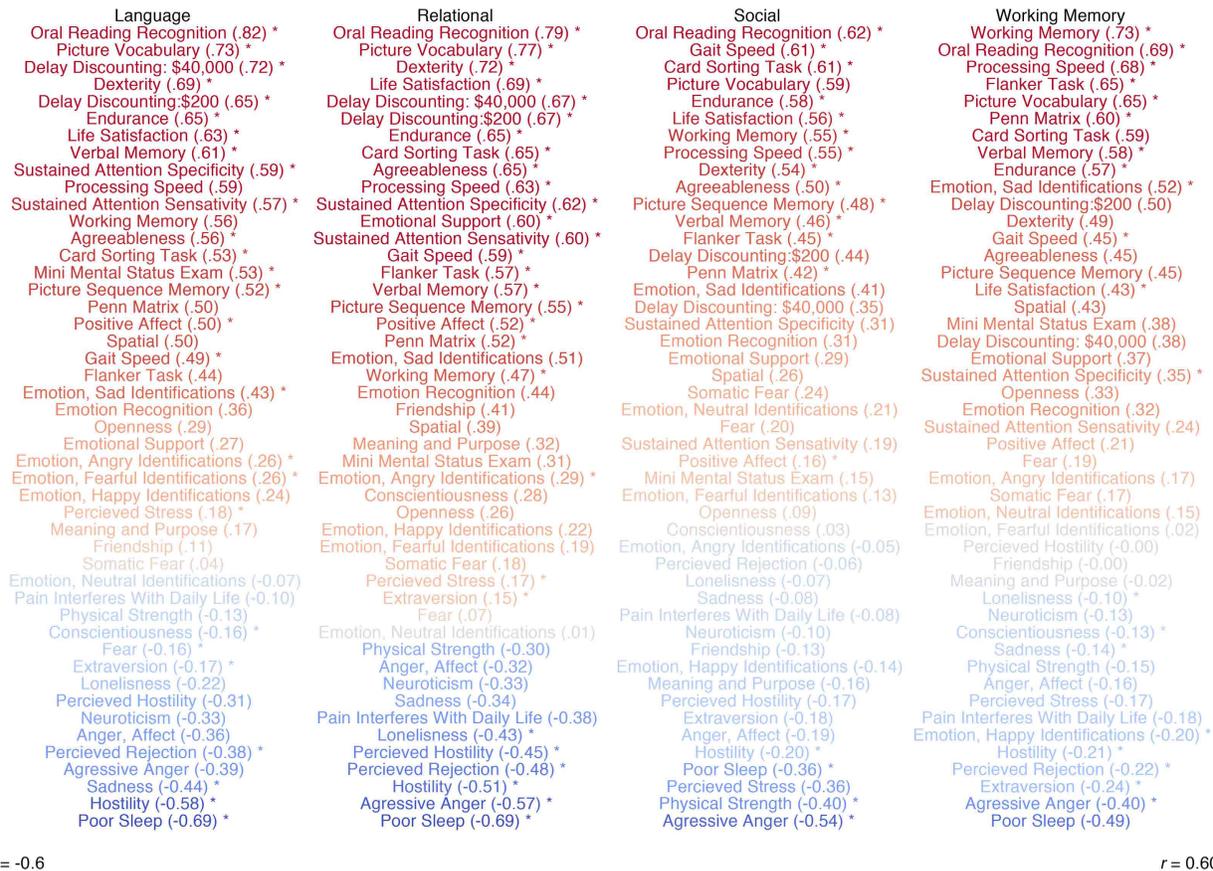

Supplementary Figure 3. **Feature correspondence between tasks and subject measures**. To test if certain hub and network structures that are optimal to each task (Language & Math, Relational, Social, or Working Memory) are also optimal for other subject measures, we measured the feature correspondence between subjects' features in the task model with the subjects' features in the same model that was fit to a given subject measure (e.g., Delayed Discounting) instead of performance in the task. The feature correspondence is shown for each task; high feature correspondence means a similar hub and network structure is optimal for the task and subject measure. Colors represent z-scored (within column) feature correspondence values. Results significant at *p*<1-e3 uncorrected, *p*<0.05 Bonferroni corrected (*N* tests= 47) are marked with an asterisk. *N* subjects = Working Memory: 473, Relational: 457,Language & Math: 471, Social: 473).

| Language | Relational | Social | Working Memory |
|---|---|---|---|
| Oral Reading Recognition (.46) * | Physical Strength (.33) * | Physical Strength (.35) * | Physical Strength (.42) * |
| Picture Vocabulary (.44) * | Oral Reading Recognition (.33) * | Penn Matrix (.35) * | Picture Vocabulary (.41) * |
| Physical Strength (.41) * | Picture Vocabulary (.31) * | Picture Vocabulary (.35) * | Oral Reading Recognition (.40) * |
| Penn Matrix (.34) * | Endurance (.28) * | Oral Reading Recognition (.31) * | Penn Matrix (.37) * |
| Spatial (.30) * | Spatial (.24) * | Spatial (.29) * | Spatial (.32) * |
| Endurance (.28) * | Openness (.24) * | Openness (.28) * | Endurance (.31) * |
| Openness (.26) * | Penn Matrix (.24) * | Endurance (.22) * | Working Memory (.29) * |
| Delay Discounting: $40,000 (.26) * | Agressive Anger (.21) * | Delay Discounting: $40,000 (.25) * | Openness (.28) * |
| Picture Sequence Memory (.24) * | Delay Discounting: $40,000 (.21) * | Delay Discounting:$200 (.23) * | Delay Discounting: $40,000 (.27) * |
| Working Memory (.23) * | Poor Sleep (.18) * | Picture Sequence Memory (.22) * | Delay Discounting:$200 (.24) * |
| Delay Discounting:$200 (.22) * | Card Sorting Task (.17) * | Verbal Memory (.22) * | Card Sorting Task (.23) * |
| Sustained Attention Specificity (.21) * | Sustained Attention Specificity (.17) * | Working Memory (.21) * | Flanker Task (.23) * |
| Agressive Anger (.21) * | Anger, Affect (.16) * | Flanker Task (.20) * | Agressive Anger (.22) * |
| Verbal Memory (.20) * | Delay Discounting:$200 (.14) | Card Sorting Task (.18) * | Gait Speed (.20) * |
| Poor Sleep (.20) * | Hostility (.14) | Agressive Anger (.18) * | Conscientiousness (.20) * |
| Life Satisfaction (.19) * | Dexterity (.14) | Dexterity (.17) * | Processing Speed (.20) * |
| Dexterity (.19) * | Picture Sequence Memory (.14) | Processing Speed (.17) * | Poor Sleep (.20) * |
| Card Sorting Task (.17) * | Flanker Task (.14) | Agreeableness (.16) * | Life Satisfaction (.20) * |
| Emotion, Angry Identifications (.17) * | Loneliness (.14) | Sustained Attention Specificity (.13) | Agreeableness (.19) * |
| Agreeableness (.15) * | Processing Speed (.13) | Poor Sleep (.12) | Sustained Attention Specificity (.18) * |
| Processing Speed (.15) * | Emotion Recognition (.12) | Life Satisfaction (.11) | Picture Sequence Memory (.18) * |
| Flanker Task (.14) | Agreeableness (.12) | Emotion Recognition (.10) | Neuroticism (.14) |
| Gait Speed (.13) | Life Satisfaction (.12) | Emotion, Neutral Identifications (.09) | Meaning and Purpose (.14) |
| Emotion Recognition (.13) | Working Memory (.12) | Fear (.09) | Verbal Memory (.13) |
| Anger, Affect (.12) | Gait Speed (.11) | Anger, Affect (.08) | Percieved Hostility (.13) |
| Loneliness (.11) | Emotion, Angry Identifications (.11) | Mini Mental Status Exam (.08) | Anger, Affect (.13) |
| Positive Affect (.11) | Emotional Support (.10) | Gait Speed (.08) | Sadness (.12) |
| Hostility (.07) | Meaning and Purpose (.09) | Sadness (.08) | Dexterity (.12) |
| Mini Mental Status Exam (.07) | Percieved Stress (.08) | Hostility (.07) | Emotion, Neutral Identifications (.10) |
| Sadness (.06) | Neuroticism (.08) | Percieved Stress (.05) | Emotion Recognition (.10) |
| Extraversion (.06) | Sadness (.07) | Emotion, Sad Identifications (.05) | Emotional Support (.09) |
| Neuroticism (.05) | Verbal Memory (.07) | Conscientiousness (.05) | Loneliness (.09) |
| Emotion, Sad Identifications (.05) | Mini Mental Status Exam (.07) | Somatic Fear (.04) | Emotion, Angry Identifications (.09) |
| Emotion, Fearful Identifications (.05) | Percieved Hostility (.05) | Emotion, Angry Identifications (.03) | Sustained Attention Sensitivity (.08) |
| Somatic Fear (.05) | Extraversion (.05) | Sustained Attention Sensitivity (.03) | Fear (.08) |
| Emotion, Neutral Identifications (.04) | Emotion, Fearful Identifications (.04) | Neuroticism (.03) | Somatic Fear (.06) |
| Conscientiousness (.04) | Positive Affect (.04) | Meaning and Purpose (.02) | Hostility (.06) |
| Fear (.03) | Conscientiousness (.04) | Emotional Support (.02) | Percieved Stress (.05) |
| Percieved Hostility (.03) | Emotion, Neutral Identifications (.03) | Emotion, Happy Identifications (.01) | Positive Affect (.04) |
| Pain Interferes With Daily Life (.03) | Emotion, Sad Identifications (.02) | Extraversion (.00) | Friendship (.04) |
| Meaning and Purpose (.02) | Sustained Attention Sensitivity (.00) | Percieved Hostility (.00) | Mini Mental Status Exam (.04) |
| Emotional Support (.02) | Pain Interferes With Daily Life (.00) | Friendship (.00) | Emotion, Sad Identifications (.03) |
| Sustained Attention Sensitivity (.01) | Percieved Rejection (.00) | Pain Interferes With Daily Life (-.01) | Emotion, Happy Identifications (.03) |
| Percieved Stress (.00) | Friendship (-.00) | Percieved Rejection (-.03) | Pain Interferes With Daily Life (-.01) |
| Friendship (-.02) | Emotion, Happy Identifications (-.02) | Emotion, Fearful Identifications (-.10) | Percieved Rejection (.00) |
| Emotion, Happy Identifications (-.04) | Somatic Fear (-.08) | | Extraversion (-.00) |
| Percieved Rejection (-.05) | Fear (-.10) | | Emotion, Fearful Identifications (-.06) |

*r* = 0.0            *r* = 0.50

Supplementary Figure 4 | **Prediction of subject measures by the model of hub diversity and locality, modularity, and network connectivity**. As in Figure 1, we predicted each subject measure using a model of hub diversity and locality, modularity, and network connectivity. This was executed for networks constructed based on the data from each task (columns). For each task and subject measure, the Pearson *r* between the real subject measure values and the predicted subject measure values are shown. Colors represent the *z*-scored (within each task) Pearson *r*. Predictions that are significant (*p*<1-e3 uncorrected, *p*<0.05 after Bonferroni correction, *N* tests = 47) are marked with an asterisk. *N*= Working Memory: 473, Relational: 457, Language & Math: 471, Social: 473).

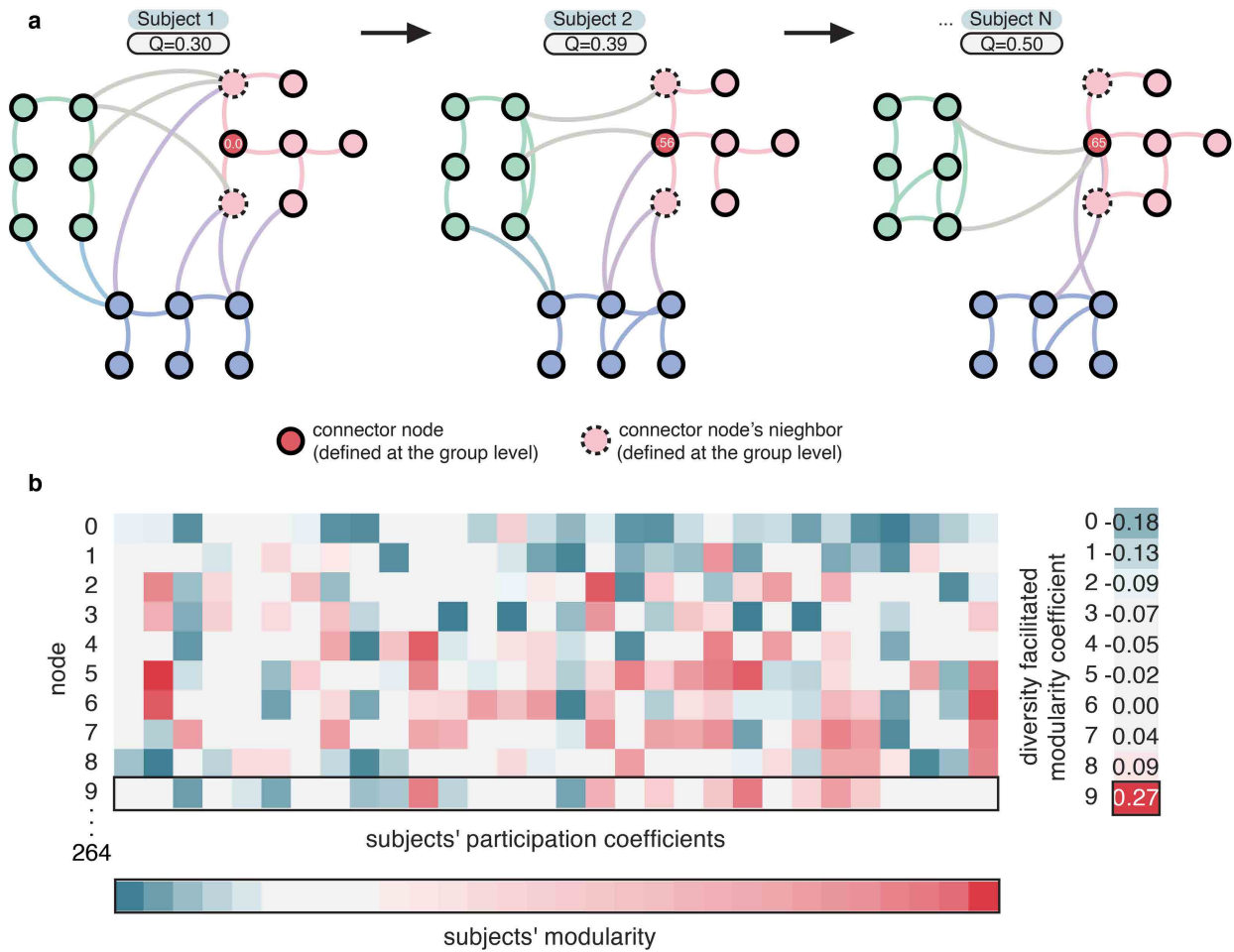

Supplementary Figure 5 | **Analyzing individual differences of hubs' diversity and locality and modularity**. When connector hubs are connected to many communities (as indicated by high participation coefficients), they can integrate information and tune connectivity optimally, allowing other regions to perform more modular local processing. Thus, across subjects, we predicted that the increased participation coefficients (i.e., diversity) of connector hubs would be correlated with preservations or increases in the modularity ($Q$) of the network. This prediction is illustrated above (**a**). Each graph represents the network structure of an individual subject. A single connector hub, shown in red, is identified based on its high mean participation coefficient across individuals; subject-level participation coefficient values are shown inside the red nodes. Local neighborhoods or *communities* are shown in green, blue, and pink, and subject-level values of modularity($Q$) are shown above each graph. Across subjects, we predict that higher participation coefficients of connector hubs will occur in more modular networks in which the neighbors of connector hubs (pink, dashed-outline) display more local connectivity. **b**, To test this prediction, for each node we measured the diversity facilitated modularity coefficient; for example, the Pearson *r* is calculated between node 9's participation coefficients across subjects (black outline) and modularity ($Q$) across subjects (black outline), resulting in a diversity facilitated modularity coefficient of 0.27 (black outline).

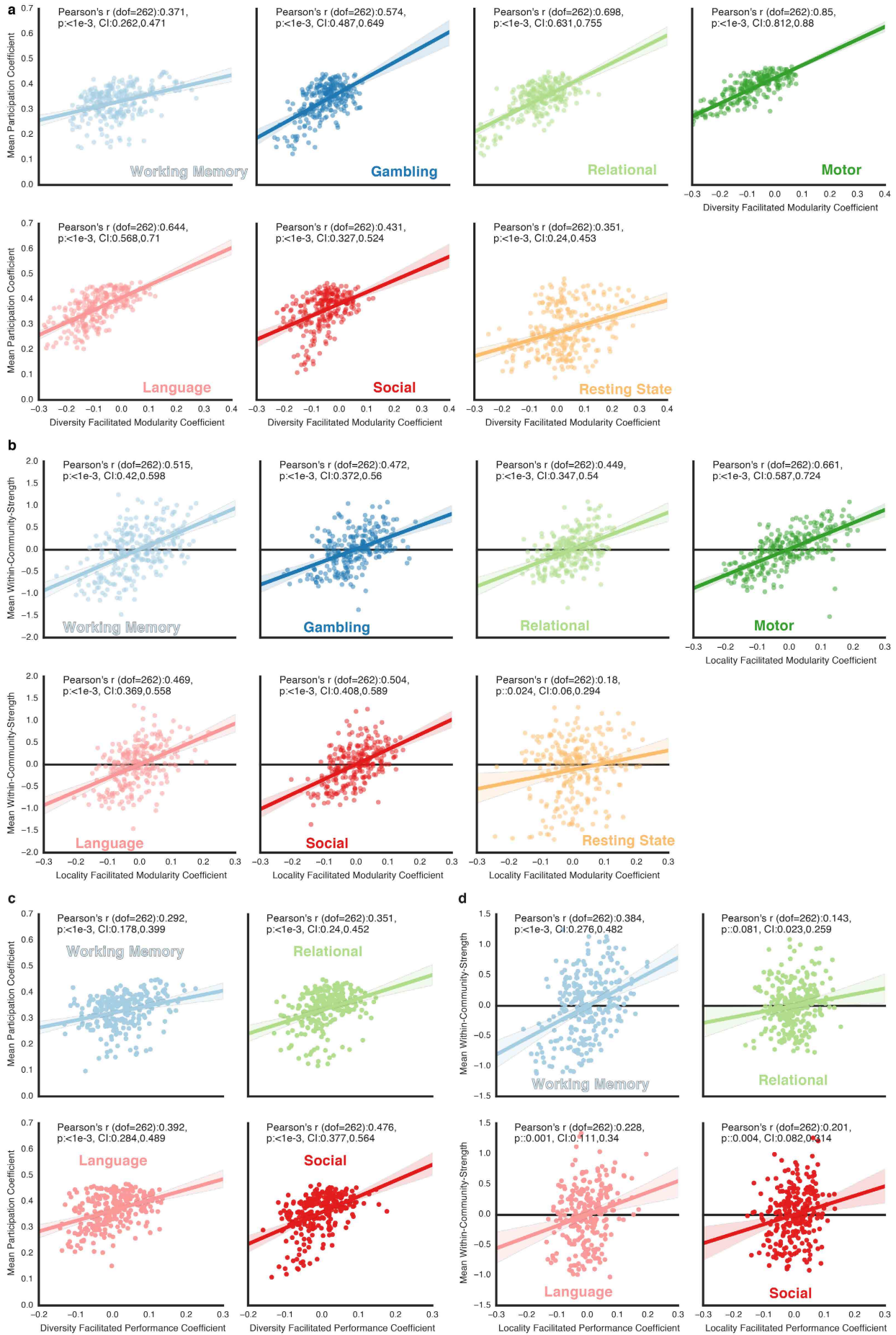

Supplementary Figure 6 | **Connector and local hubs facilitate network modularity and task performance**. **a**, correlations between a node's mean participation coefficient (across subjects) and the node's diversity facilitated modularity coefficient (the Pearson *r* correlation coefficient between that node's participation coefficients and modularity values ($Q$) across subjects). Correlations were calculated for each cognitive state. Each dot represents a node in the brain's functional network. Shaded areas represent 95 percent confidence intervals. In all cognitive states, there was a significant positive correlation between a node's mean participation coefficient and the node's diversity facilitated modularity coefficient. This demonstrates that connector hubs facilitate increased modularity. **b**, correlations between a node's mean within community strength (across subjects) and the node's locality facilitated modularity coefficient (Pearson *r* correlation coefficient between that node's within community strengths and modularity values ($Q$) across subjects). In all cognitive states, there was a significant positive correlation between a node's within community strength and the node's locality facilitated modularity coefficient. This demonstrates that local hubs facilitate increased modularity. **c**, correlations between a node's mean participation coefficient (across subjects) and the node's diversity facilitated performance coefficient (the Pearson *r* correlation coefficient between that node's participation coefficients and task performance across subjects). Correlations were calculated for each cognitive state. Each dot represents a node in the brain's functional network. Shaded areas represent 95 percent confidence intervals. In all cognitive states, there was a significant positive correlation between a node's mean participation coefficient and the node's diversity-facilitated performance coefficient. This demonstrates that connector hubs facilitate increased task performance. **d**, correlations between a node's mean within community strength (across subjects) and the node's locality facilitated performance coefficient (Pearson *r* correlation coefficient between that node's within community strengths and task performance across subjects). In all cognitive states (except Relational, Bonferroni (*n* tests=4) *p*=0.081, uncorrected *p*=0.02), there was a significant positive correlation between a node's within community strength and the node's locality facilitated performance coefficient. Bonferroni corrected *p* values are shown (*n* tests=7(a,b), and 4(c,d). *N*=264, the number of nodes in the graph. This demonstrates that local hubs facilitate increased task performance.

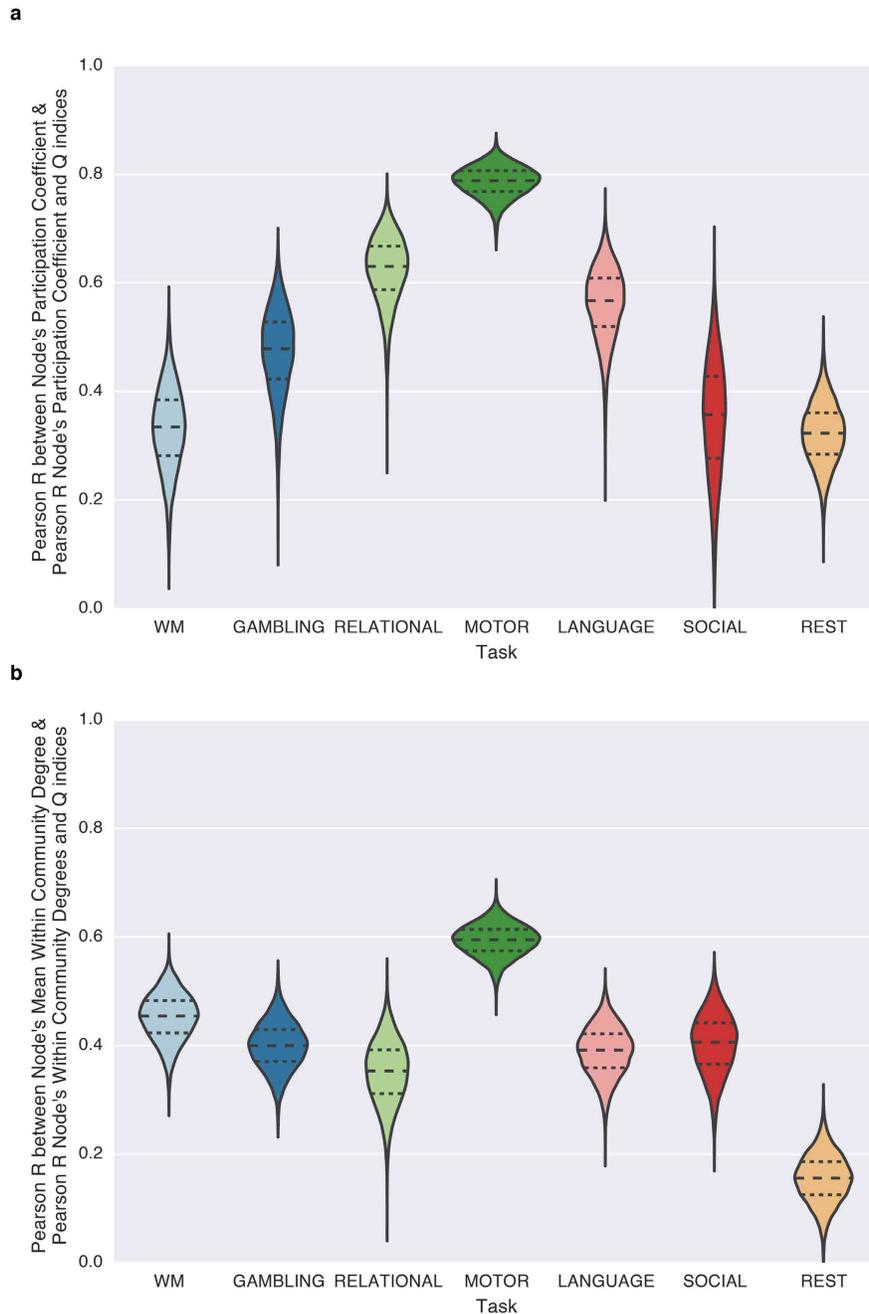

Supplementary Figure 7 | **Cross validation of diversity and locality facilitated modularity coefficients**. Connector hubs and local hubs are defined by a higher participation coefficient and within community strength, respectively, across subjects on average. To avoid potential dependencies, we estimated the mean within community strength or participation coefficient of each node in one half of subjects, and the correlation between a nodes' within community strength or participation coefficients and $Q$ indices—the node's locality and diversity facilitated modularity coefficients—in the other half, testing 10,000 split of subjects. On average, correlations were still significant in every state. **a**, The distribution of Pearson $r$ values between a node's mean participation coefficient (defined in half the subjects) and the diversity facilitated modularity coefficient (defined in the other half of the subjects). **b**, The distribution of Pearson $r$ values between a node's mean within community strength (defined in half the subjects) and the locality facilitated modularity coefficient (defined in the other half of the subjects). $N$=264, the number of nodes in the graph, for each $r$ value calculation, (dof=262). The mean and quartiles are marked in each violin.

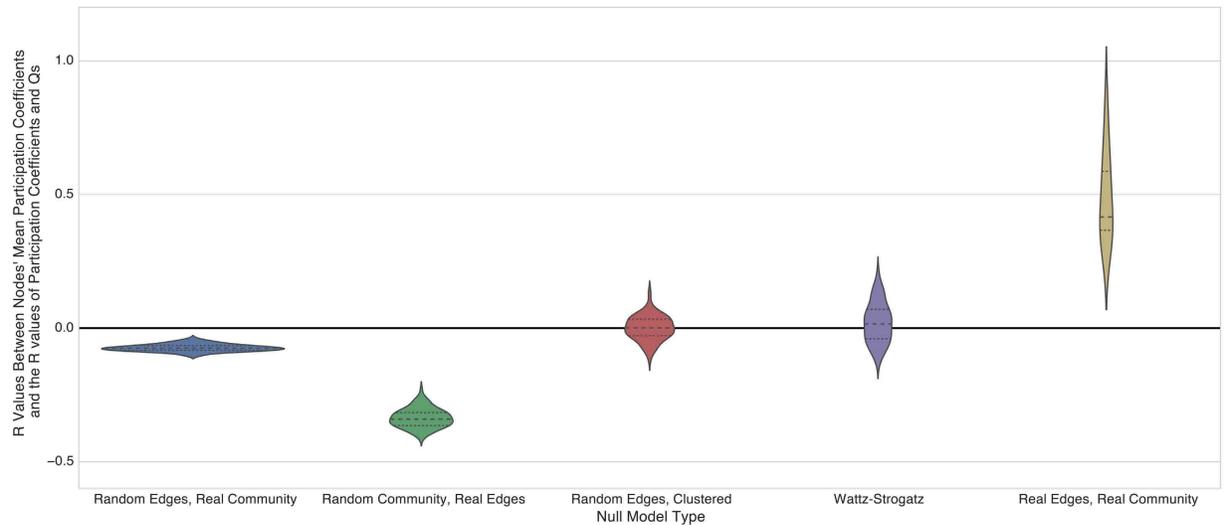

Supplementary Figure 8 | **Null Models**. We tested four null models of diversity facilitated modularity coefficients. (1) Random Edges, Real Community utilizes the true partition of nodes into communities that was uncovered by the application of community detection to each subject's intact resting-state graph (thresholded at a weighted graph density of 0.10, chosen based on this cost being the median cost from our original analyses), but we randomly permuted the edges uniformly after the partition was identified. (2) Random Community, Real Edges utilizes the partition of nodes into communities that was uncovered by the application of community detection to each subject's intact resting-state graph (thresholded at a weighted graph density of 0.10), but we then randomly permuted the assignment of nodes to communities (this retains the same number of communities and sizes of communities as the true partition). The edges remain in their true locations. (3) Random Edges, Clustered permutes the edges in each subject's resting-state graph uniformly at random. We then applied community detection to this permuted graph to identify a partition of nodes into communities, and then calculated the participation coefficient of each node based on that partition. We used a cost of 0.05, as denser random graphs result in just one community, which results in trivial participation coefficients of 0. For each instantiation ($n$=100) of each of these three models, we generated a graph for each subject using the subject's original graph and that particular null model. Finally, (4) we generated Wattz-Strogatz small world graphs. Each graph had 264 nodes with each node initially connected to its 7 neighbors in the lattice. We set the rewiring probability to 0.25. This results in $Q$ values of roughly 0.40 and a binary density of roughly 0.05. 100 graphs were generated for each instantiation($n$=100) of the model. For each null model, we then calculated the diversity facilitated modularity coefficient. Violin plots are shown for the distribution of the correlation between each node's mean participation coefficient and each node's diversity facilitated modularity coefficient across 100 instantiations of each model. For comparison, the distribution from the original analysis across tasks (Real Edges, Real Community) is also shown. $N$=264, the number of nodes in the graph, dof=262.

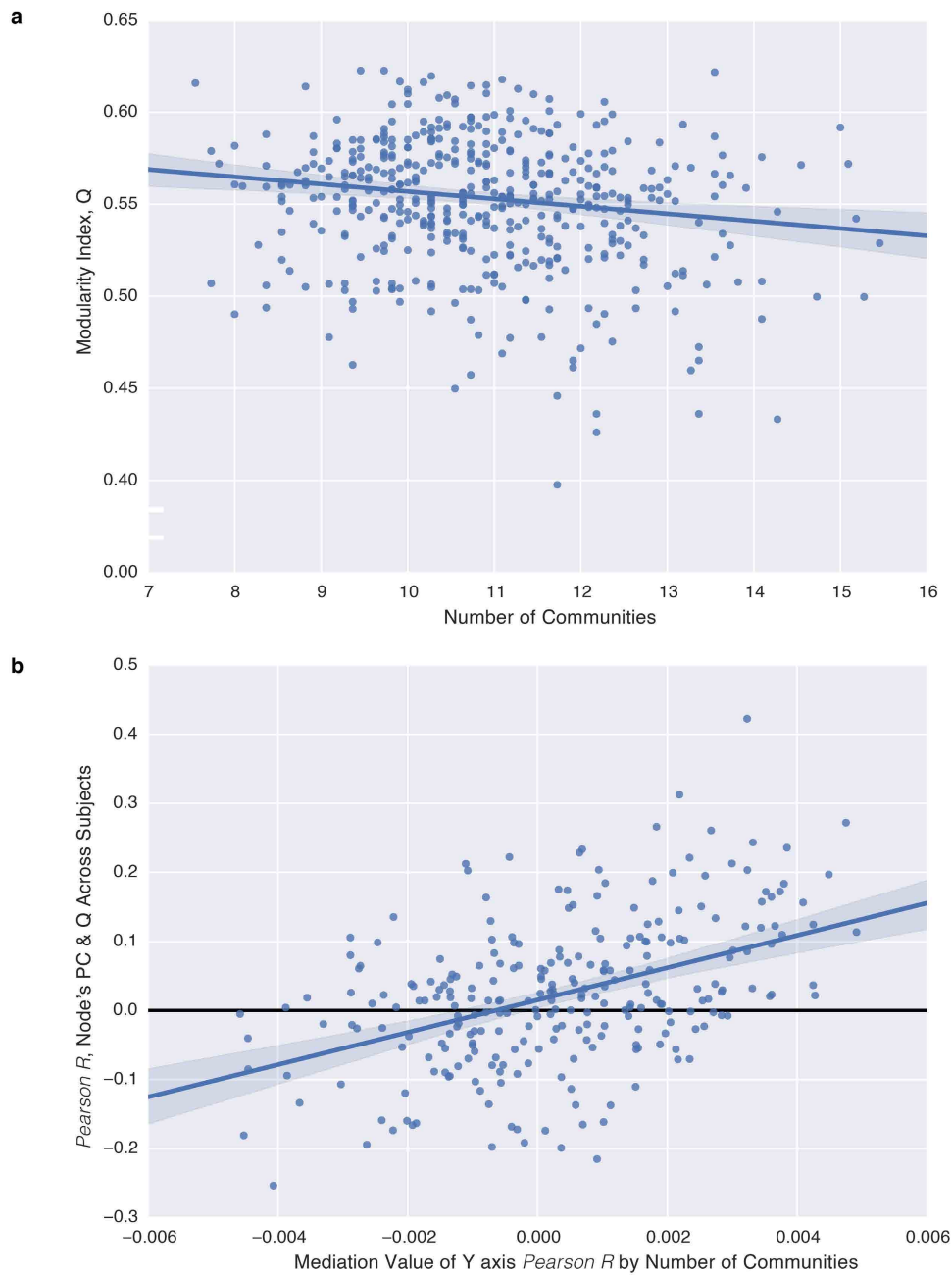

Supplementary Figure 9 | **Participation coefficients, the number of communities in the network, and modularity ($Q$).** **a**, Negative correlation between the number of communities in subjects' graphs (averaged across graph densities in the range 0.05-0.15) and the $Q$ value of the graphs (also averaged across graph densities in the range 0.05-0.15). **b**, We performed a mediation analysis between each node's participation coefficients, the number of communities in the graphs, and the $Q$ values of the graphs, with the number of communities being the mediator. Mediation values are plotted for each node on the x-axis. The y-axis is the Pearson $r$ between each node's participation coefficients and $Q$ indices across subjects' graphs—the diversity facilitated modularity coefficient. Shaded areas represent a 95 percent confidence interval. $N$=264, the number of nodes in the graph, dof=262.

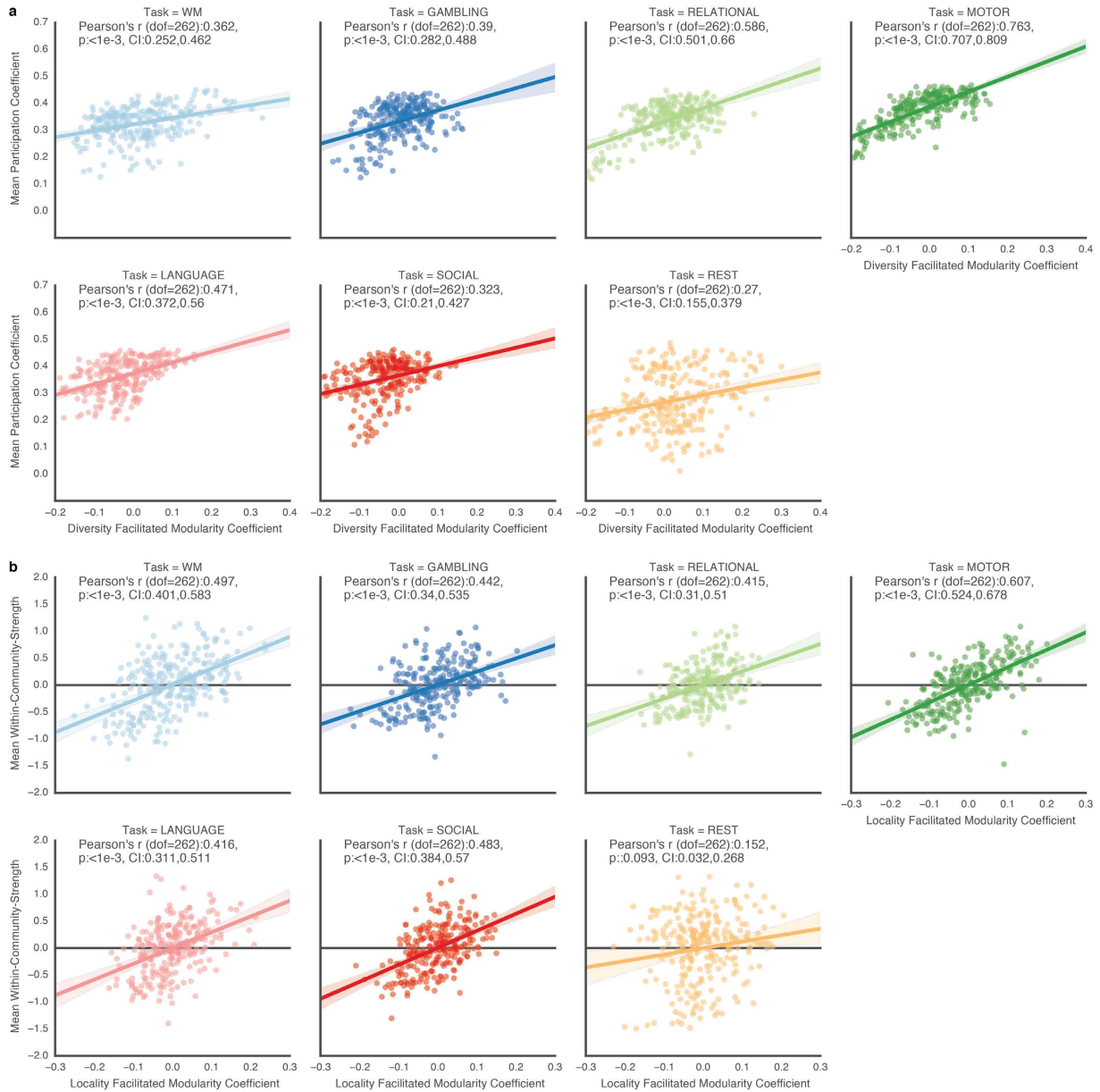

Supplementary Figure 10 | **Diversity and locality facilitated modularity coefficients controlling for the number of communities**. **a**, Correlations between a node's mean participation coefficient (across subjects) and the node's diversity facilitated modularity coefficient (the Pearson $r$ correlation coefficient between that node's participation coefficients and modularity values ($Q$) across subjects after regressing out the number of communities in the network from $Q$). Correlations were calculated for each cognitive state. Each dot represents a node in the brain's functional network. Shaded areas represent 95 percent confidence intervals. In all cognitive states, there was a significant positive correlation between a node's mean participation coefficient and the node's diversity-facilitated modularity coefficient. **b**, Correlations between a node's mean within community strength (across subjects) and the node's locality facilitated modularity coefficient (Pearson $r$ correlation coefficient between that node's within community strength and modularity values ($Q$) across subjects after regressing out the number of communities in the network from $Q$. In all cognitive states (except Resting state (uncorrected $p$=0.0132, Bonferroni corrected $p$=0.093), there was a significant positive correlation between a node's within community strength and the node's locality facilitated modularity coefficient. All $p$ values are Bonferroni corrected ($n$ tests=7). $N$=264, the number of nodes in the graph. $N$ subjects=Working Memory:475, Gambling:473, Relational: 458, Motor:475, Language:472, Social: 474, Rest: 476.

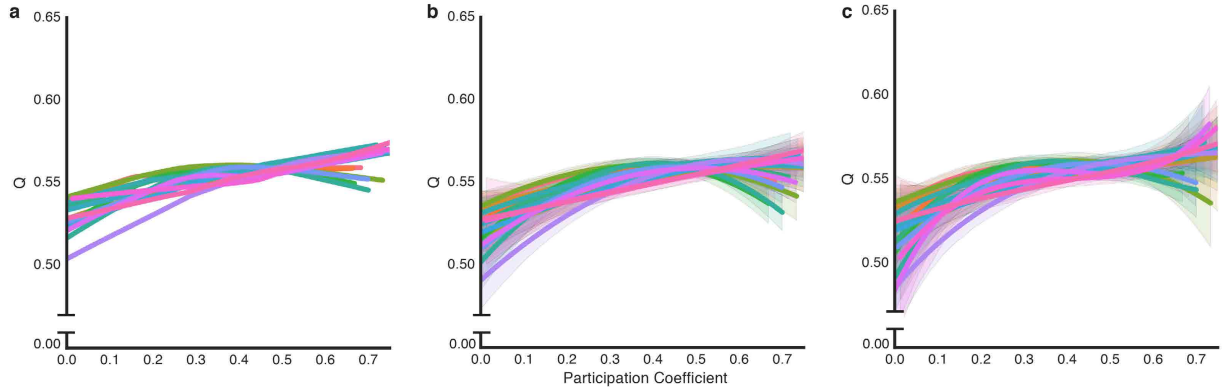

Supplementary Figure 11 | **Relationships between individual connector hubs' participation coefficients and *Q***. To test if the relationship between a connector hub's participation coefficients and *Q*—the connector hubs' diversity facilitated modularity coefficient—was linear, we fit three regression models to individual connector hubs' participation coefficients and *Q* values across subjects. In each plot, each line is the relationship between a single connector hub's participation coefficients and *Q* values across subjects. Only nodes with positive Pearson *r* values are shown. **a**, Locally weighted scatter-plot smoother fit. **b**, 2nd order fit. **c**, 3rd order fit. The relation between many connector hubs' participation coefficients and *Q* indices across subjects is well captured by a first order fit, with the hub's maximal participation coefficients (0.7; the mathematically upper limit is 0.9$\bar{9}$) corresponding to maximal *Q*. However, for example, some connector hubs' participation coefficients correspond to the maximal *Q* at a participation coefficient of 0.4-0.5, and then *Q* decreases at higher participation coefficients of that connector hub. Shaded areas represent 95 percent confidence intervals. *N*=476.

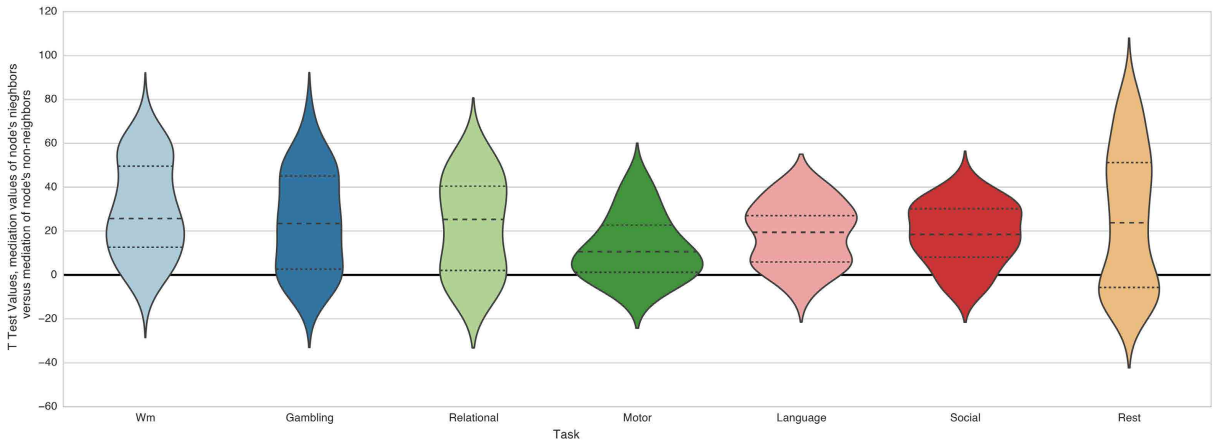

Supplementary Figure 12 | **Connector hubs' mediation of neighboring nodes' edges**. To investigate if the relationship between a connector hub's participation coefficient and *Q* is mediated primarily by that connector hub's neighbors' edge pattern increasing *Q* (per our prediction), we executed, for each connector hub node (*i*), a *t*-test between the absolute mediation values of node (*i*)'s neighbors' edges versus the absolute mediation values of node (*i*)'s non-neighbors' edges (neighbors were defined based on edges present between the two nodes in a graph at a density of 0.15, which was used because it is the densest density we utilized in our analyses; neighbors' and non-neighbors' edges connecting to the node (*i*) were ignored). Mediation values were based on the edge mediating between node (*i*)'s participation coefficients and modularity values (*Q*); the distributions of mediation values was tested for normality using D'Agostino and Pearson's omnibus test $k^2$. All distributions were confirmed as normal ($k^2$>100.0, *p<0.001* for all tasks). The distribution of *t*-values for connector hubs is shown for each task. The mean quartiles are marked. In general, connector hubs showed higher mediation values with edges of its neighbors compared to the edges of its non-neighbors (i.e., *t*>0). Moreover, these *t*-values were significantly higher than the same *t*-values calculated with local hubs (*t*(dof:1358): 3.892, *p*:0.0001, Cohen's *d*:0.219, CI:1.887,6.62).

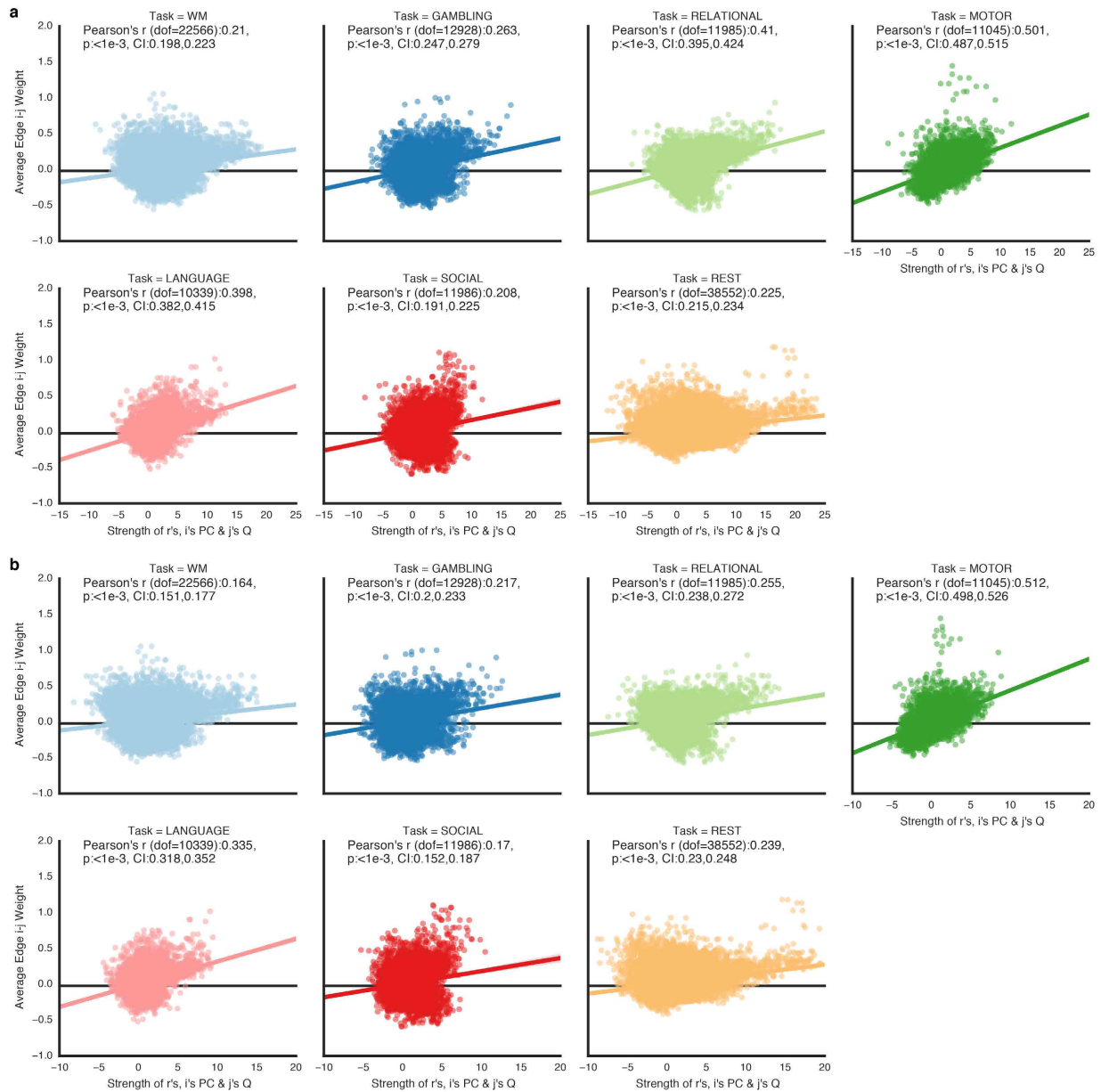

Supplementary Figure 13 | **Connector hubs' relationships with individual edges' weights**. **a**,**b**, For each task, for each pair of nodes *i* and *j*, we calculated the correlation between (*x*) how strongly node *i* increased the within community edge strength of node *j* (the sum of Pearson r values between the participation coefficients of node *i* and the within community edges of node *j* minus the sum of Pearson r values between the participation coefficients of node *i* and the between community edges of node *j*) with (y) the average connectivity edge weight between node *i* and node *j*. **a** Includes all positive edges, while **b** includes the strongest 25 percent of edges. Bonferroni corrected *p* values are shown in all plots (*n* tests = 7). These results show that connector hubs are predominately tuning the connectivity of their neighbors. Shaded areas represent 95 percent confidence intervals. *N*=dof+2 for each panel.

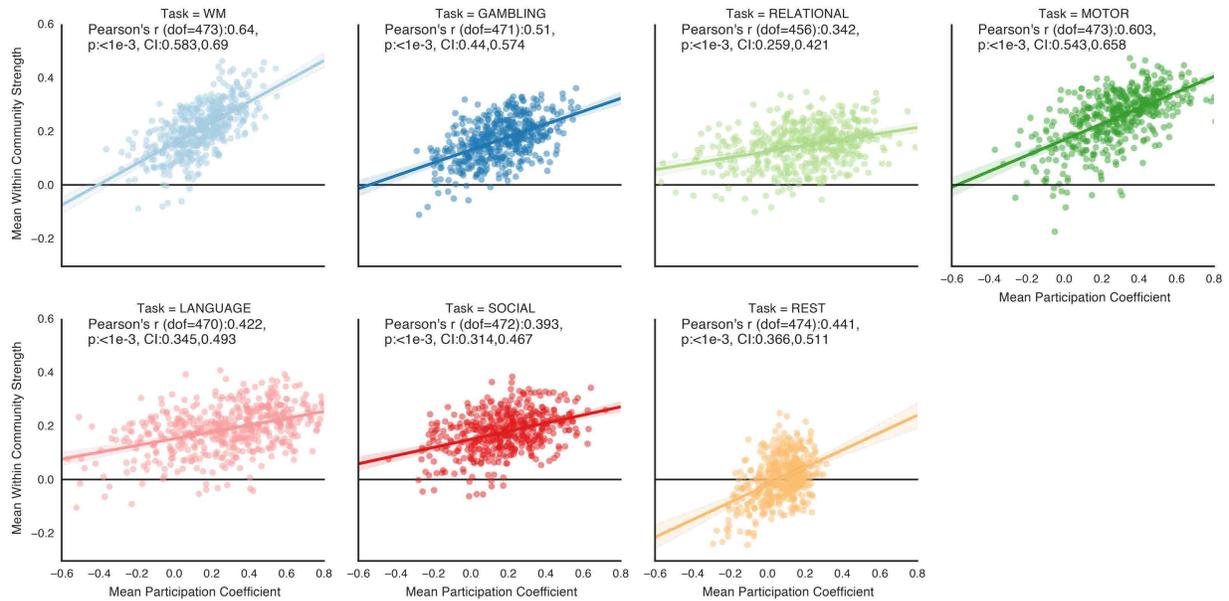

Supplementary Figure 14 | **Connector hubs and local hubs are interrelated**. For each task, across subjects, the correlation between the subject's connector hubs' mean participation coefficient and the subject's local hubs' mean within community strengths. In every task, if a subject's connector hubs were diverse, local hubs were local. Here, connector and local hubs were defined based on nodes for which their diversity and locality (respectively) facilitated modularity coefficients were positive. For this calculation, participation coefficients were z-scored within each subject, as within community strengths are z-scored within each subject. Bonferroni corrected *p* values are reported in all plots(*n* tests = 7). Shaded areas represent 95 percent confidence intervals. *N*=Working Memory:475, Gambling:473, Relational: 458, Motor:475, Language:472, Social: 474, Rest: 476.

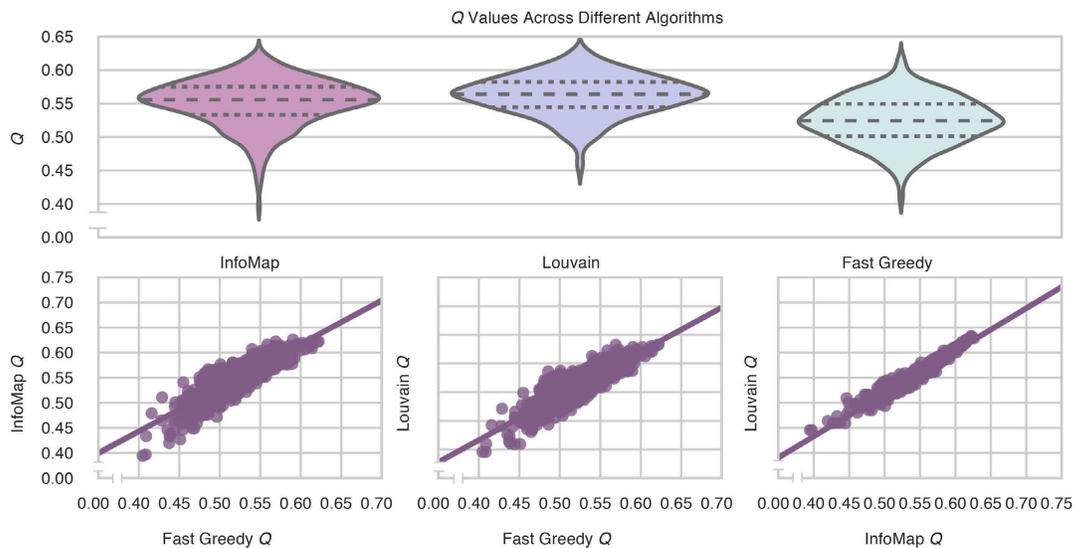

Supplementary Figure 15 | **Modularity quality indices from three community detection algorithms**. We used *Q* indices repeatedly in our analyses. However, the community detection algorithm that we utilized, InfoMap, does not explicitly maximize *Q*. To see if this could potentially impact our analyses, we compared *Q* values from InfoMap to two popular algorithms, Fast-Greedy and Louvain, that explicitly maximize *Q*. The mean *Q* value, as in our analyses, was taken across costs of 0.05 to 0.15. **a**, Distribution, across subjects, of *Q* values of each algorithm. **b**, Correlation between Q values across algorithms. Shaded areas represent 95 percent confidence intervals. In all panels, *N*=476, dof=474.

1. Bertolero, M. A., Yeo, B. T. T., & D'Esposito, M. The modular and integrative functional architecture of the human brain. *Proc. Natl. Acad. Sci. U.S.A.* **112,** E6798–807 (2015).
2. Bertolero, M. A., Yeo, B. T. T., & D'Esposito, M. The diverse club. *Nature Communications* **8,** 1277 (2017).
3. Guimerà, R., Guimera, R., Amaral, L. A. N. & Amaral, L. Functional cartography of complex metabolic networks. *Nature* **433,** 895–900 (2005).
4. Meunier, D., Lambiotte, R. & Bullmore, E. T. Modular and hierarchically modular organization of brain networks. *Front. Neurosci.* **4,** 200 (2010).
5. Bertolero, M. A., Yeo, B. T. T., & D'Esposito, M. The diverse club. *Nature Communications* **8,** 1–10 (2017).
6. Bassett, D. S. *et al.* Efficient Physical Embedding of Topologically Complex Information Processing Networks in Brains and Computer Circuits. *PLoS Comput Biol* **6,** e1000748 (2010).
7. Laughlin, S. B., de Ruyter van Steveninck, R. R. & Anderson, J. C. The metabolic cost of neural information. *Nature Neuroscience* **1,** 36–41 (1998).
8. Lord, L.-D., Expert, P., Huckins, J. F. & Turkheimer, F. E. Cerebral Energy Metabolism and the Brain's Functional Network Architecture: An Integrative Review. *Journal of Cerebral Blood Flow & Metabolism* **33,** 1347–1354 (2013).
9. Raichle, M. E. & Gusnard, D. A. Appraising the brain's energy budget. *Proceedings of the National Academy of Sciences* **99,** 10237–10239 (2002).
10. Harris, J. J. & Attwell, D. The Energetics of CNS White Matter. *Journal of Neuroscience* **32,** 356–371 (2012).
11. Kuzawa, C. W. *et al.* Metabolic costs and evolutionary implications of human brain development. *Proceedings of the National Academy of Sciences* **111,** 13010–13015 (2014).
12. Krienen, F. M., Yeo, B. T. T., Ge, T., Buckner, R. L. & Sherwood, C. C. Transcriptional profiles of supragranular-enriched genes associate with corticocortical network architecture in the human brain. *Proc. Natl. Acad. Sci. U.S.A.* **113,** E469–78 (2016).
13. Hawrylycz, M. *et al.* Canonical genetic signatures of the adult human brain. *Nature Neuroscience* **18,** 1832–1844 (2015).
14. Wagner, G. P. & Zhang, J. The pleiotropic structure of the genotype-phenotype map: the evolvability of complex organisms. *Nature Reviews Genetics* **12,** 204–213 (2011).
15. Clune, J., Mouret, J.-B. & Lipson, H. The evolutionary origins of modularity. *Proc. Biol. Sci.* **280,** 1–9 (2013).
16. Kashtan, N., Kashtan, N., Alon, U. & Alon, U. Spontaneous evolution of modularity and network motifs. **102,** 13773–13778 (2005).
17. Simon, H. A. in *Facets of Systems Science* 457–476 (Springer US, 1991). doi:10.1007/978-1-4899-0718-9_31
18. Fodor, J. & Fodor, J. *The Modularity of Mind*. (MIT Press, 1983).
19. Coltheart, M. Modularity and cognition. *Trends in Cognitive Sciences* **3,** 115–120 (1999).

20. Robinson, P. A., Henderson, J. A., Matar, E., Riley, P. & Gray, R. T. Dynamical Reconnection and Stability Constraints on Cortical Network Architecture. *Physical Review Letters* **103,** 108104 (2009).
21. Clune, J., Mouret, J.-B. & Lipson, H. The evolutionary origins of modularity. *Proc. Biol. Sci.* **280,** 20122863–20122863 (2013).
22. Tosh, C. R., Tosh, C. R., McNally, L. & McNally, L. The relative efficiency of modular and non-modular networks of different size. *Proc. Biol. Sci.* **282,** 20142568–20142568 (2015).
23. Stevens, A. A., Tappon, S. C., Garg, A. & Fair, D. A. Functional brain network modularity captures inter- and intra-individual variation in working memory capacity. *PLOS ONE* **7,** e30468 (2012).
24. Arnemann, K. L. *et al.* Functional brain network modularity predicts response to cognitive training after brain injury. *Neurology* **84,** 1568–1574 (2015).
25. Modular Brain Network Organization Predicts Response to Cognitive Training in Older Adults. *PLoS ONE* (2016).
26. Warren, D. E. *et al.* Network measures predict neuropsychological outcome after brain injury. *Proc. Natl. Acad. Sci. U.S.A.* **111,** 14247–14252 (2014).
27. Gratton, C., Nomura, E. M., Pérez, F. & D'Esposito, M. Focal Brain Lesions to Critical Locations Cause Widespread Disruption of the Modular Organization of the Brain. *Journal of Cognitive Neuroscience* **24,** 1275–1285 (2012).
28. van den Heuvel, M. P. & Sporns, O. Network hubs in the human brain. *Trends in Cognitive Sciences* **17,** 683–696 (2013).
29. Rubinov, M., Ypma, R. J. F., Watson, C. & Bullmore, E. T. Wiring cost and topological participation of the mouse brain connectome. *Proc. Natl. Acad. Sci. U.S.A.* **112,** 10032–10037 (2015).
30. Ferreira, F. R. M., Nogueira, M. I. & Defelipe, J. The influence of James and Darwin on Cajal and his research into the neuron theory and evolution of the nervous system. *Front Neuroanat* **8,** 1 (2014).
31. Guimera, R., Mossa, S., Turtschi, A. & Amaral, L. A. N. The worldwide air transportation network: Anomalous centrality, community structure, and cities' global roles. *Proceedings of the National Academy of Sciences* **102,** 7794–7799 (2005).
32. Guimerà, R., Sales-Pardo, M. & Amaral, L. A. N. Classes of complex networks defined by role-to-role connectivity profiles. *Nature Physics* **3,** 63–69 (2007).
33. Power, J. D., Schlaggar, B. L., Lessov-Schlaggar, C. N. & Petersen, S. E. Evidence for Hubs in Human Functional Brain Networks. **79,** 798–813 (2013).
34. van den Heuvel, M. P. & Sporns, O. An Anatomical Substrate for Integration among Functional Networks in Human Cortex. **33,** 14489–14500 (2013).
35. Scholtens, L. H., Schmidt, R., de Reus, M. A. & van den Heuvel, M. P. Linking Macroscale Graph Analytical Organization to Microscale Neuroarchitectonics in the Macaque Connectome. **34,** 12192–12205 (2014).
36. Cole, M. W. *et al.* Multi-task connectivity reveals flexible hubs for adaptive task control. *Nature Neuroscience* **16,** 1348–1355 (2013).

37. Yeo, B. T. T. *et al.* Functional Specialization and Flexibility in Human Association Cortex. *Cereb Cortex* **25,** 3654–3672 (2015).
38. Bassett, D. S., Yang, M., Wymbs, N. F. & Grafton, S. T. Learning-induced autonomy of sensorimotor systems. *Nature Neuroscience* **18,** 744–751 (2015).
39. Spadone, S. *et al.* Dynamic reorganization of human resting-state networks during visuospatial attention. *Proc. Natl. Acad. Sci. U.S.A.* **112,** 8112–8117 (2015).
40. Gratton, C., Laumann, T. O., Gordon, E. M., Adeyemo, B. & Petersen, S. E. Evidence for Two Independent Factors that Modify Brain Networks to Meet Task Goals. *Cell Reports* **17,** 1276–1288 (2016).
41. Yamashita, M., Kawato, M. & Imamizu, H. Predicting learning plateau of working memory from whole-brain intrinsic network connectivity patterns. *Scientific Reports* **5,** 7622 (2015).
42. Baldassarre, A. *et al.* Individual variability in functional connectivity predicts performance of a perceptual task. *Proc. Natl. Acad. Sci. U.S.A.* **109,** 3516–3521 (2012).
43. Sala-Llonch, R. *et al.* Brain connectivity during resting state and subsequent working memory task predicts behavioural performance. *Cortex* **48,** 1187–1196 (2012).
44. Sadaghiani, S., Poline, J.-B., Kleinschmidt, A. & D'Esposito, M. Ongoing dynamics in large-scale functional connectivity predict perception. *Proc. Natl. Acad. Sci. U.S.A.* **112,** 8463–8468 (2015).
45. Shine, J. M. *et al.* The Dynamics of Functional Brain Networks: Integrated Network States during Cognitive Task Performance. *Neuron* **92,** 544–554 (2016).
46. Lerman, C. *et al.* Large-scale brain network coupling predicts acute nicotine abstinence effects on craving and cognitive function. *JAMA Psychiatry* **71,** 523–530 (2014).
47. Finn, E. S. *et al.* Functional connectome fingerprinting: identifying individuals using patterns of brain connectivity. *Nature Neuroscience* **18,** 1664–1671 (2015).
48. Van Essen, D. C. *et al.* The WU-Minn Human Connectome Project: an overview. *NeuroImage* **80,** 62–79 (2013).
49. Power, J. D. *et al.* Functional Network Organization of the Human Brain. *Neuron* **72,** 665–678 (2011).
50. Smith, S. M. *et al.* A positive-negative mode of population covariation links brain connectivity, demographics and behavior. *Nature Neuroscience* **18,** 1565–1567 (2015).
51. Bullmore, E. & Sporns, O. The economy of brain network organization. *Nat Rev Neurosci* **74,** 47–14 (2012).
52. Cox, R. W. AFNI: software for analysis and visualization of functional magnetic resonance neuroimages. *Comput. Biomed. Res.* **29,** 162–173 (1996).
53. Cole, M. W., Bassett, D. S., Power, J. D., Braver, T. S. & Petersen, S. E. Intrinsic and Task-Evoked Network Architectures of the Human Brain. **83,** 238–251 (2014).
54. Siegel, J. S., Mitra, A., Laumann, T. O. & Seitzman, B. A. Data quality influences observed links between functional connectivity and behavior. *Cerebral …* (2016).

55. Gu, S., Satterthwaite, T. D. & Medaglia, J. D. Emergence of system roles in normative neurodevelopment. in (2015).
56. Zalesky, A., Fornito, A. & Bullmore, E. On the use of correlation as a measure of network connectivity. *NeuroImage* **60,** 2096–2106 (2012).
57. Rosvall, M., Rosvall, M., Bergstrom, C. T. & Bergstrom, C. T. Maps of random walks on complex networks reveal community structure. **105,** 1118–1123 (2008).
58. Brandes, U., Delling, D. & Gaertler, M. On modularity clustering. *IEEE transactions on …* (2008).
59. Lancichinetti, A. & Fortunato, S. Community detection algorithms: A comparative analysis. *Phys. Rev. E* **80,** 056117–11 (2009).
60. Berenstein, A. J., Piñero, J., Furlong, L. I. & Chernomoretz, A. Mining the Modular Structure of Protein Interaction Networks. *PLOS ONE* **10,** e0122477 (2015).
61. Clauset, A., Newman, M. & Moore, C. Finding community structure in very large networks. *Phys. Rev. E* (2004).
62. Blondel, V. D., Guillaume, J.-L., Lambiotte, R. & Lefebvre, E. Fast unfolding of communities in large networks. *Journal of Statistical Mechanics: Theory and Experiment* **2008,** P10008 (2008).
63. WU-Minn HCP 1200 Subjects Data Release Reference Manual. 1–213 (2018). https://www.humanconnectome.org/storage/app/media/documentation/s1200/HCP_S1200_Release_Reference_Manual.pdf
64. Shen, X. *et al.* Using connectome-based predictive modeling to predict individual behavior from brain connectivity. *Nat Protoc* **12,** 506–518 (2017).